\documentclass[12pt]{article}

\usepackage[utf8]{inputenc}
\usepackage[T1]{fontenc}

\usepackage{setspace} 
\usepackage[vmargin = 1.5in, hmargin =1.5in]{geometry}

\usepackage[usenames,dvipsnames,svgnames]{xcolor}
\usepackage{amsmath, amssymb, amsfonts, graphicx, tikz,pdflscape, mathtools, amsthm, upgreek, bm,pgfplots,csvsimple,multirow, multicol, booktabs}
\usepackage{verbatim}
\usepackage{natbib}
\usepackage{accents}
\newcommand{\ubar}[1]{\underaccent{\bar}{#1}}
\usepackage{needspace}
\def\citeapos#1{\citeauthor{#1}'s (\citeyear{#1})}
\usepackage{comment}
\usepackage[inline]{enumitem}
\usepackage{tocvsec2}
\usepackage{threeparttable}
\usetikzlibrary{calc, cd}
\usepackage[skip = 10pt]{caption}
\allowdisplaybreaks[1]
\allowdisplaybreaks[1]

\definecolor{Blue}{RGB}{86,180,233}
\definecolor{Orange}{RGB}{230,159,0}
\definecolor{Green}{RGB}{0,158,115}
\definecolor{GmailBlue}{RGB}{42, 93, 176} 
\usepackage[
	pagebackref,
	colorlinks=true,
	citecolor= GmailBlue,
	linkcolor=GmailBlue,
	urlcolor = GmailBlue
]{hyperref}

\newcommand{\bibtexorder}[1]{}

\usepackage{pgfplots}
\usepgfplotslibrary{groupplots,colorbrewer}
\pgfplotsset{compat=newest}
\pgfplotsset{cycle list/Set1}
\usepackage{tikz}
\usetikzlibrary{matrix,calc,shapes,arrows.meta,positioning}
\tikzset{
    vertex/.style = {shape=circle,draw, minimum size = 1.8em, inner sep = 0pt},
    edge/.style = {->,> = latex}
}

\usepackage[capitalize,noabbrev]{cleveref}

\newtheoremstyle{break}
{}
{}
{\itshape}
{}
{\bfseries}
{}
{\newline}
{}

\theoremstyle{break}
\newtheorem{thm}{Theorem}
\newtheorem*{theorem*}{Theorem}
\newtheorem*{cor*}{Corollary}

\newtheorem{prop}{Proposition}
\newtheorem{lem}{Lemma}

\crefname{prop}{Proposition}{Propositions}
\crefname{thm}{Theorem}{Theorems}
\crefname{lem}{Lemma}{Lemmas}
\crefname{blem}{Lemma}{Lemmas}

\theoremstyle{definition}
\newtheorem{defn}{Definition}
\newtheorem{exmp}{Example}

\newtheorem{rem}[thm]{Remark}
\newtheorem*{rem*}{Remark}
\newtheorem*{claim*}{Claim}

\def\a{\alpha}

\def\d{\delta}
\def\e{\varepsilon}

\def\th{\theta}

\def\l{\lambda}

\def\s{\sigma}

\def\x{\chi}

\def\D{\Delta}
\def\Th{\Theta}
\def\L{\Lambda}

\def\R{\mathbf{R}}

\def\BB{\mathcal{B}}

\def\EE{\mathcal{E}}

\def\KK{\mathcal{K}}

\def\XX{\mathcal{X}}
\def\YY{\mathcal{Y}}

\def\P{\mathbf{P}}

\DeclareMathOperator{\E}{\mathbf{E}}
\DeclareMathOperator{\supp}{supp} 
\DeclareMathOperator*{\argmax}{argmax}

\DeclareMathOperator{\ST}{st}

\newcommand{\Abs}[1]{\left\lvert #1 \right\rvert}

\newcommand{\Paren}[1]{\left( #1 \right)}

\newcommand{\Brac}[1]{\left[ #1 \right]}

\newcommand{\Set}[1]{\left\{ #1 \right\}}

\newcommand{\de}{\mathop{}\!\mathrm{d}}

\usepackage{datetime}
\newdateformat{specialdate}{\THEDAY~\monthname[\THEMONTH] \THEYEAR}

\title{Probabilistic Verification in Mechanism Design\thanks{%
We thank our former advisors, Dirk Bergemann and Stephan Lauermann, for continual guidance. For useful feedback, we thank audiences at Aalto, Arizona, Bonn, Brown, Columbia, Princeton, PSE, Harvard/MIT, Humboldt,  Rochester, SAET, and Yale. For helpful discussions, we thank Tilman B\"{o}rgers, Marina Halac, Johannes H\"{o}rner, Navin Kartik, Andreas Kleiner, Daniel Kr\"{a}hmer, Bart Lipman, Benny Moldovanu, Stephen Morris, Jacopo Perego, Larry Samuelson, Sebastian Schweighofer-Kodritsch, Roberto Serrano, Philipp Strack, Roland Strausz, Juuso V\"{a}lim\"{a}ki, and Alex Wolitzky. An extended abstract for this paper appeared in EC '19. We are grateful for the opportunity to attend the 28th Jerusalem Summer School in Economic Theory, which planted the first seeds of this project. Funding from the German Research Foundation (DFG) through CRC TR 224 (Project B04) is gratefully acknowledged.
}}

\author{%
	Ian Ball%
	\thanks{Department of Economics, MIT, ianball@mit.edu.}
	\and 
	Deniz Kattwinkel%
	\thanks{Department of Economics, UCL, denizkattwinkel@gmail.com.}
}
\date{\specialdate\today}

\begin{document}

\maketitle

\begin{abstract}
We introduce a model of probabilistic verification in mechanism design. The principal elicits a message from the agent and then selects a test to give the agent. The agent's true type determines the probability with which he can pass each test. We characterize whether each type has an associated test that best screens out all other types. If this condition holds, then the testing technology can be represented in a tractable reduced form. We use this reduced form to solve for profit-maximizing mechanisms with verification. As the verification technology varies, the solution continuously interpolates between the no-verification solution and full surplus extraction.
\end{abstract}

\noindent \emph{Keywords:} probabilistic verification, ordering tests, evidence.

\noindent \emph{JEL Codes:} D82, D86.

\newpage
\onehalfspacing

\section{Introduction} \label{sec:intro}

In the standard paradigm of mechanism design, the principal elicits information from the agents, but the  principal cannot \emph{verify} whether the agents are being truthful. In many applications, however, claims about private information can be verified. Sellers have long offered discounts to certain groups such as students, seniors, or veterans, and sellers increasingly make targeted offers to different consumer segments. To verify a buyer's eligibility for an exclusive offer, many sellers use identity verification platforms such as ID.me and SheerID. If a buyer claims to be eligible for a particular offer, he is directed to a portal, which asks identifying questions or requests documentation, such as a student ID or a company email address. Depending on the buyer's responses, the platform's proprietary algorithm either accepts or rejects the buyer's claim. In other contexts, governments verify income reports to determine eligibility for means-tested programs. Insurers verify the legitimacy of insurance claims. 
None of these verification systems is perfect---false claims sometimes go undetected. In this sense, verification is \emph{probabilistic}. 

The goal of this paper is to introduce a tractable model of probabilistic verification. A parsimonious model of probabilistic verification, directly generalizing \citeapos{GreenLaffont1986} deterministic model, would specify for any types $\th$ and $\th'$ the probability $\a(\th' |\th)$ with which type $\th$ can ``pass'' as type $\th'$.\footnote{For example, \cite{CaragiannisEtal2012} and \cite{FerraioliVentre2018} take this approach.} Call this function $\a$ the \emph{authentication rate}. The difficulty is that with unrestricted communication, the authentication rate is generally endogenous. Whether type $\th$ can ``pass'' as type $\th'$ depends on what the principal demands of an agent who claims to be of type $\th'$, e.g., which questions must be answered correctly or which documents must be provided. 

We model \emph{probabilistic verification} by endowing the principal with a set of pass--fail tests.  A test could be a particular set of questions or a request for certain documentation. We represent a test by its type-dependent passage rate: type $\th$ can pass test $\tau$ with probability $\pi (\tau | \th)$. We assume that every type can intentionally fail any test. For example, the agent could leave a question blank (whether or not he knows the answer) or decline to provide the requested documentation (whether or not he has it).

The principal chooses how to utilize the testing technology within a mechanism. Formally, we consider the following protocol. The principal elicits a type report from the agent. Based on the report, the principal selects one test to give the agent. The agent sees the test and privately chooses whether to try on the test.  This choice is costless. If the agent tries, then his passage probability depends on the test and his type, according to the function $\pi$.  If the agent does not try, then he fails the test with certainty. The principal observes whether the agent passes or fails---but not whether the agent tried---and then makes a decision.

Our analysis proceeds in two parts---methodology and then applications. 

In the first part of our analysis, we study whether there exists a canonical assignment of a test to each type. For each type $\th$, we introduce an associated order on tests. Intuitively, test $\tau$ is \emph{more $\th$-discerning} than test $\psi$ if type $\th$ performs relatively better on test $\tau$ than on test $\psi$, compared to \emph{every} other type. The formal definition requires that there is a ``conversion'' from $\tau$-scores to $\psi$-scores that is fair for type $\th$ but disadvantageous for all other types. This score conversion is similar to a Blackwell garbling of an experiment, but our order neither implies nor is implied by Blackwell's order. 

We use our order on tests to simplify the principal's implementation problem. Consider two tests, $\tau$ and $\psi$, such that $\tau$ is more $\th$-discerning than $\psi$. \cref{res:replacement} says that any social choice function that the principal can implement by giving test $\psi$ to type $\th$ can also be implemented by giving test $\tau$ to type $\th$. We apply this logic repeatedly to obtain \cref{res:implementation}: If each type $\th$ has an associated test that is most $\th$-discerning, then there is no loss in assuming that the principal gives each type the associated test. In a contemporaneous paper, \cite{BenPorathDekelLipmanWP} prove a related result in their model of stochastic evidence acquisition; we compare our model with theirs in \cref{sec:discussion}.

If each type $\th$ has an associated most-$\th$ discerning test, then the testing technology can be represented by a single authentication rate: type $\th$ is authenticated as type $\th'$ if and only if he passes the most $\th'$-discerning test. The principal's design problem reduces to a tractable optimization problem  involving this authentication rate. An authentication rate that represents a testing technology in this way is called \emph{most-discerning}. Not all authentication rates are most-discerning. We characterize the class of most-discerning authentication rates. Our condition generalizes the conditions imposed in the literature on deterministic verification \citep{GreenLaffont1986} and evidence \citep{LipmanSeppi1995,BullWatson2007}.  

For the second part of our analysis, we turn to applications, taking a most-discerning authentication rate as a primitive. Unlike in models of \emph{deterministic} verification, we can use the Myersonian local approach. Consider a seller who can imperfectly verify a potential buyer's membership in different market segments. It is more difficult for the seller to distinguish buyers who are in market segments with closer valuations. If the seller has a single indivisible good, it is no longer optimal for the seller to post one price. The seller prefers to charge higher prices to higher-valuation market segments. If a buyer in a higher segment claims to be in a lower segment, there is a chance that he is authenticated and charged a lower price. But there is also a chance that his misreport is detected and he does not receive the good. Under the optimal price schedule, these deviations are unprofitable. 

To solve for profit-maximizing mechanisms in general quasilinear settings with verification, we derive a new expression for the virtual value that reflects the verification technology. As verification ranges from uninformative to perfectly informative, the virtual value increases from the classical virtual value to the true valuation.  The associated revenue-maximizing mechanism continuously interpolates between the classical, no-verification solution and full surplus extraction.

The rest of the paper is organized as follows.  \cref{sec:model} presents our model of testing.  \cref{sec:model_discussion} discusses our modeling choices. \cref{sec:order} introduces the discernment orders and characterizes whether a single testing function suffices for all implementation. \cref{sec:reduced_form} characterizes the class of most-discerning authentication rates. \cref{sec:applications} solves for revenue-maximizing mechanisms. \cref{sec:nonbinary} extends the model to allow for nonbinary tests. \cref{sec:discussion} discusses related literature. 
The conclusion is \cref{sec:conclusion}. The main proofs are in \cref{sec:main_proofs}. Additional results are in  \cref{sec:additional_results}.

\section{Model} \label{sec:model}

We model probabilistic verification by endowing the principal with a testing technology. The principal can commit to use this technology (and communicate with the agent) however she wishes. 

\subsection{Setting} \label{sec:setting}

\paragraph{Principal--agent environment}
There are two players: a principal (she) and an agent (he). 
The agent has a private type $\th \in \Th$, drawn from a commonly known distribution. The principal controls a decision $x\in X$.%
\footnote{Transfers could be one component of the decision $x$.} The agent and the principal have bounded, type-dependent utilities $u(x,\th)$ and $v(x, \th)$, respectively. We extend these functions linearly to $\D (X) \times \Th$. 

\paragraph{Verification}
There is a testing technology $(T, \pi)$, which consists of a set $T$ of pass--fail tests and a passage rate
\[
    \pi \colon T \times \Th \to [0,1], 
\]
where $\pi (\tau | \th)$ denotes the probability with which type $\th$ can pass test $\tau$. The spaces $X$, $\Th$, and $T$ are assumed to be Polish spaces.\footnote{We further assume that the primitives $u$, $v$, and $\pi$ are Borel measurable and that mechanisms are universally measurable. The details are in  \cref{sec:measurability}.} 

The principal can give the agent one test from the set $T$.\footnote{If the principal can give multiple tests, then the resulting compound test can be included in $T$. A compound test may have more scores than ``pass'' and ``fail.''  \cref{sec:nonbinary} extends the model to allow for nonbinary tests.} The agent observes the selected test and chooses whether to \emph{try} on the test. This choice is costless. If the agent tries, his passage probability is determined by $\pi$. If the agent does not try, then he fails with certainty. The principal observes the test score (``pass'' or ``fail''), but not whether the agent tried. Thus, there is moral hazard as well as adverse selection. In \cref{sec:model_discussion}, we discuss why this modeling choice is natural in many applications. \cref{sec:nonbinary} considers tests with more than two scores. 

\paragraph{Mechanisms and strategies}

\begin{figure}
\begin{center}
\begin{tikzpicture}
	\draw [->, > = latex, thick] (-2,0) -- (11,0);
	\foreach \x in {0,2,4,6,10}
	\draw [thick] (\x cm,3pt) -- (\x cm,-3pt);
	
	\node [shape=circle, draw, minimum size = 1.5em, inner sep = 0pt, fill = white] (N1) at (-2,0) {$N$};
	\draw [->, > = latex] (N1) -- (-2, 0.75) node [anchor = south] {$\theta$};
	
	\node [shape=circle, draw, minimum size = 1.5em, inner sep = 0pt, fill = white] (N2) at (8,0) {$N$};
	\draw [->, > = latex] (N2) -- (8, -0.75) node [anchor = north] {$1$ or $0$};
	
	\node [anchor = south west, align = left] at (-0.75, 0.5) {Agent \\sends\\ message};
	\node [anchor = north west, align = left] at (1.25, -0.5) {Principal \\ selects \\ test};
	\node [anchor = north west, align = left] at (3.25, -0.5) {Principal \\ sends \\ message};
	\node [anchor = south west, align = left] at (5.25, 0.5) {Agent \\ chooses to \\ try or not };
	\node [anchor = north west, align = left] at (9.25, -0.5) {Principal  \\ makes \\ decision};
\end{tikzpicture}
\end{center}
\caption{Timing}
\label{fig:timing}
\end{figure}

The principal can commit to an arbitrary dynamic mechanism. We consider protocols of the following form, shown in \cref{fig:timing}. First the principal elicits a message from the agent. Based on the message, the principal selects a test and then sends a message to the agent. The agent sees the realized test and the message and then privately chooses whether to try on the test.  Nature draws the test score: ``pass'' (denoted $1$) or ``fail'' (denoted $0$). The principal observes this score---but not whether the agent tried---and then makes a decision.

Formally, a mechanism is a tuple $(M, M'; t, r', g)$ consisting of message spaces $M$ and $M'$ for the two rounds of messaging, a testing rule $t \colon M \to \D (T)$, a messaging rule $r' \colon M \times T \to \D(M')$, and an outcome rule $g \colon M \times T \times M' \times \{0,1\} \to \D(X)$. A strategy for the agent is a pair $(r,a)$ consisting of a messaging strategy $r \colon \Th \to \D(M)$ and an action strategy $a \colon \Th \times M \times T \times M' \to [0,1]$, which specifies the probability with which the agent tries on the test he is given.

\subsection{Implementation}

We introduce two social choice objects. A \emph{social choice function} is a map from $\Th$ to
$\D (X)$, which specifies a decision lottery for each type. To keep track of which test is given, we define an \emph{extended social choice function} to be a map from $\Th$ to $\D (T \times X)$, which specifies for each type a joint lottery over tests and decisions.
A mechanism and a strategy together \emph{implement} an (extended) social choice function $f$ if (i) the strategy is a best response to the mechanism and (ii) the composition of the mechanism and the strategy induces $f$. An (extended) social choice function is \emph{implementable} if there exist a mechanism and a strategy that implement it.

We show below that it is without loss to focus on a special class of direct mechanisms that induce the agent to (i) report his type truthfully and (ii) try on whichever test he is given. In these mechanisms, the principal's message to the agent is omitted. Formally, a \emph{canonical mechanism} is a pair $(t,g)$ consisting of a testing rule $t \colon \Th \to \D(T)$ and an outcome rule $g \colon \Th \times T \times \{0,1\} \to \D(X)$, which specifies a decision lottery as a function of the reported type, the test given to the agent, and the agent's score on that 
test. Given such a mechanism, a strategy for the agent is a pair $(r,a)$ consisting of a reporting strategy $r \colon \Th \to \D(\Th)$ and an action strategy $a \colon \Th \times \Th \times T \to [0,1]$, which specifies the probability with which the agent tries as a function of his true type, his reported type, and the test. An (extended) social choice function $f$ is \emph{canonically implementable} if $f$ is implemented by some canonical mechanism $(t,g)$ and some strategy $(r,a)$ in which
$r$ is the identity and $a (\th, \th, \tau) = 1$ for all types $\th$ and all tests $\tau$ in $\supp t(\th)$. In this case, we say that $(t,g)$ \emph{canonically implements} $f$.

\begin{prop}[Revelation principle] \label{res:revelation_principle}
Every implementable (extended) social choice function is canonically implementable. 
\end{prop}

The proof has two parts. First, a standard argument \citep[see][]{Myerson1982} shows that every implementable social choice function can be implemented by a truthful and obedient mechanism. The second part is specific to testing. Consider a truthful, obedient mechanism. Whenever the principal recommends the agent to not try on a test, we modify the mechanism as follows. The principal recommends that the agent try on the test. Then, if the agent passes that test, the principal selects the decision as if the agent had failed. Now passing and failing result in the same decision. The agent is willing to follow the recommendation, and the resulting outcome is unchanged. Since every type can fail every test, this modification of the mechanism introduces no new deviation outcomes. Since the principal always recommends that the agent try, the principal's message conveys no information and hence can be dropped.

\section{Discussion of the model} \label{sec:model_discussion}

We discuss two important features of the model: the agent's choice of whether to try on the test, and the principal's choice of a testing rule. 

\subsection{Trying on the test} 
\label{sec:trying}

When the principal gives the agent a test, the agent privately chooses whether to try or intentionally fail. If the agent fails, the principal cannot observe whether the failure was intentional. This assumption is reasonable in our motivating applications.\footnote{In their model of adaptive testing, \cite{DebStewart2018} make the same assumption about the agent's performance on each ``task.'' The principal commits to an adaptive sequence of binary tasks and then assigns a final verdict---pass or fail. Each agent type wants to pass. In our model, by contrast, the principal chooses from a richer space of decisions, and different agent types have different preferences over those decisions.}
If the test asks the agent a question and the agent leaves the question blank, then the principal cannot tell whether the agent knows the answer. On an aptitude test, both high- and low-ability types are able to perform poorly.\footnote{\citet[p.~74]{Myerson1984} gives the example of playing the piano. A good pianist can intentionally play poorly, but a bad pianist cannot play well.} If the agent does perform poorly, the principal cannot tell whether the agent is capable of performing well. Finally, if the test requests a document and the agent does not provide it, then the principal cannot tell whether the agent has the document. Indeed, our model nests previous models of deterministic hard evidence. In those models, the agent's choice to present evidence that he possesses is an ``inalienable action'' \citep[p.~76]{BullWatson2007}.
 
The importance of allowing intentional failure is illustrated in the following example.

\begin{exmp}[Passing v. failing] \label{ex:skipping}
Consider the problem of allocating a single desirable good to an agent with two possible types, $\th_1$ and $\th_2$. There is a single test. If the agent tries on this test, then type $\th_1$ passes with certainty and type $\th_2$ fails with certainty. In order to allocate the good to type $\th_1$ only,
the principal can give the agent the good if and only if he passes the test. On the other hand, the principal cannot allocate the good to type $\th_2$ only. If the principal gives the good to the agent if and only if he fails the test, then type $\th_1$ would intentionally fail in order to receive the good. 
\end{exmp}

The logic of \cref{ex:skipping} holds more generally. Without intentional failing,  ``pass'' and ``fail'' would be arbitrary, interchangeable labels. On any test, if type $\th_1$ is more likely to pass than type $\th_2$, then type $\th_2$ is more likely to \emph{fail} than type $\th_1$. Thus, without intentional failing, each test unavoidably links the ability of type $\th_1$ to mimic $\th_2$ with the ability of type $\th_2$ to mimic $\th_1$.

Finally, we discuss two alternative assumptions about the agent's control over the test result: (i) \emph{observable skipping} and (ii) \emph{exogenous scores}. Under (i), the agent cannot intentionally fail a test, but he can ``skip'' the test; skipping is observed by the principal. Under (ii), the agent can neither intentionally fail nor skip a test. These assumptions are nested in terms of the power afforded to the principal. Every social choice function that is implementable under our model is also implementable under (i),\footnote{Given a canonical mechanism that is truthful and obedient in our model, the induced social choice function can be replicated under (i) by treating ``skip'' as ``fail'' on each test.} and every social choice function implementable under (i) is also implementable under (ii).\footnote{Under (i), the analogue of \cref{res:revelation_principle} still holds; the argument is essentially the same as in our model, with ``skipping'' in place of ``intentionally failing.'' Under canonical implementation, the agent never skips a test, so the incentive constraints are preserved if the agent's option to skip a test is removed.} In \cref{ex:skipping}, allocating the good only to type $\th_2$ is implementable under both specifications (i) and (ii).\footnote{The same social choice functions are implementable under (i) and (ii) whenever there is a decision that all types consider to be the worst (since this worst decision can be used to punish skipping). This condition is called TIWO for ``type-independent worst option'' in \citeapos{Strausz2016wp} model of \emph{deterministic} evidence. For an example of a social choice function that is implementable under (ii), but not under (i), reinterpret their Example 1 \citep[p.~16]{Strausz2016wp}.}

\subsection{Test choice}

In order to cleanly separate verification from communication,  we explicitly model the principal's choice of a test as part of the mechanism. As a result, the effective authentication rate faced by the agent is endogenous. Treating the authentication rate as an exogenous primitive can introduce difficulties, as illustrated in the following example, adapted from \cite{GreenLaffont1986}. 

\begin{figure}
\begin{center}
\begin{tikzpicture}
     \node [vertex] (1) at (0,0)  {$\theta_1$};
	    \node [vertex, fill = Blue!50] (1) at (0,0)  {$\theta_1$};
	 
	 \node [vertex] (2) at (1.25,2) {$\theta_2$};
	 \node [vertex, fill = Blue!50] (2) at (1.25,2) {$\theta_2$};
	 
     \node [vertex] (3) at (2.5,0) {$\theta_3$};
	  \node [vertex] (3) at (2.5,0)  {$\theta_3$};
	 
	 \draw [thick, edge, in=165, out=195,loop] (1) to (1);
	 
	 \draw [thick, edge, in=75, out=105,loop, Blue] (2) to (2);

	  \draw [thick, edge, in=-15,out=15,loop] (3) to (3);
	 \draw [thick, edge, in=-15,out=15,loop] (3) to (3);
	 
	 \draw [thick, edge] (1) to (2);
	 \draw [thick, edge, Blue] (1) to (2);
	 
    \draw [thick, edge] (2) to (3);
	\draw [thick, edge] (2) to (3);
	 
	 \draw [thick, edge] (3) to (1);
	 
\end{tikzpicture}
\hspace{2cm}
\begin{tikzpicture}
     \node [vertex] (1) at (0,0)  {$\theta_1$};
	    \node [vertex] (1) at (0,0)  {$\theta_1$};
	 
	 \node [vertex] (2) at (1.25,2) {$\theta_2$};
	 \node [vertex, fill = Blue!50] (2) at (1.25,2) {$\theta_2$};
	 
     \node [vertex] (3) at (2.5,0) {$\theta_3$};
	  \node [vertex, fill = Blue!50] (3) at (2.5,0)  {$\theta_3$};
	 
	 \draw [thick, edge, in=165, out=195,loop] (1) to (1);
	 
	 \draw [thick, edge, in=75, out=105,loop] (2) to (2);
	 \draw [thick, edge, in=75, out=105,loop] (2) to (2);

	  \draw [thick, edge, in=-15,out=15,loop] (3) to (3);
	 \draw [thick, edge, in=-15,out=15,loop, Blue] (3) to (3);
	 
	 \draw [thick, edge] (1) to (2);
	 \draw [thick, edge] (1) to (2);
	 
    \draw [thick, edge] (2) to (3);
	\draw [thick, edge, Blue] (2) to (3);
	 
	 \draw [thick, edge] (3) to (1);
	 
\end{tikzpicture}
\end{center}
\caption{Directed graph representing a $\{0,1\}$-valued authentication rate}
\label{fig:challenge}
\end{figure}

\begin{exmp}[Exogenous authentication rate] \label{sec:cautionary-example}
There is a single agent with three possible types, denoted $\th_1, \th_2, \th_3$. In \cref{fig:challenge}, the directed graph (shown twice) represents the verification technology: there is an edge from $\th$ to $\th'$ if type $\th$ can ``pass'' as type $\th'$. The principal decides whether to allocate a good to the agent. Every type wants the good.
 
Each copy of the graph illustrates a social choice function. On the left, this function allocates the good to types $\th_1$ and $\th_2$ (which are shaded). This cannot be implemented by giving the good to the agent if and only if he passes as type $\th_1$ or as type $\th_2$. Then type $\th_3$ would pass as type $\th_1$ to get the good as well. Instead, the principal must give the good to the agent if and only if he passes as type $\th_2$. Types $\th_1$ and $\th_2$ can do so, but type $\th_3$ cannot. On the right, the social choice function allocates the good to types $\th_2$ and $\th_3$ (which are shaded). Symmetrically, this can be implemented only by giving the good to the agent if and only if he passes as type $\th_3$.

According to the directed graph, type $\th_1$ can ``pass'' as type $\th_2$. But type $\th_1$ can copy type $\th_2$'s equilibrium strategy only in the equilibrium of the left mechanism (where type $\th_2$ passes as type $\th_2$), but not in the right mechanism (where type $\th_2$ passes as type $\th_3$). 
\end{exmp}

As \cref{sec:cautionary-example} illustrates, the authentication rate $\a (\cdot | \cdot)$ implicitly (a) introduces a family of tests, and (b) assigns to each type $\th$ a test, so that ``passing'' as type $\th$ means passing the test assigned to type $\th$. There is no guarantee, however, that this is the ``right'' assignment of tests to types. In \cref{sec:cautionary-example},  type $\th_2$ must be given different tests in order to implement different allocation rules. We model this test choice as part of the principal's protocol. Moreover, our model allows for an unrestricted test set (possibly larger than the type space), unrestricted communication, and test randomization.\footnote{Test randomization is useful if different tests are needed to deter deviations by different types; see \cref{sec:mixing_randomization}  for an example.}

\section{Ordering tests} \label{sec:order}

In this section, we introduce a family of orders on tests. We use these orders to identify a smaller class of testing rules that suffices for all implementation.

\subsection{Discernment orders}

For a fixed type $\th$, our order captures whether one test is better than another at distinguishing type $\th$ from all other types. 

\begin{defn}[$\th$-discernment] \label{def:discernment_order} 
Fix a type $\th$. Test $\tau$ is \emph{more $\th$-discerning} than test $\psi$, denoted $\tau \succeq_\th \psi$, if there exist probabilities $k_1$ and $k_0$ with $k_1 \geq k_0$ such that
\begin{enumerate}[label = (\roman*), ref = \roman*]
    \item \label{it:theta_equality}  $\pi (\tau | \th) k_1 + (1 - \pi (\tau | \th)) k_0 = \pi (\psi | \th)$;
    \item \label{it:theta_prime_inequality} $\pi (\tau | \th') k_1 + (1 - \pi (\tau | \th')) k_0 \leq \pi (\psi | \th')$ for all types $\th'$ with $\th' \neq \th$.
\end{enumerate}
\end{defn}
The interpretation is that after the agent takes test $\tau$, his score $s_\tau \in  \{0,1\}$ can be converted into a score $s_\psi \in \{0,1\}$ according to the transition probabilities $\P ( s_\psi = 1 | s_\tau = 1) = k_1$ and $\P ( s_\psi = 1 | s_\tau = 0) = k_0$. The inequality $k_1 \geq k_0$ ensures that passing (rather than failing) test $\tau$ weakly increases the converted pass probability. Condition \ref{it:theta_equality} says that this score conversion is fair for type $\th$. If type $\th$ tries on test $\tau$ and his score is converted, he is just as likely to pass as if he tries on test $\psi$ directly. Condition \ref{it:theta_prime_inequality} says that this score conversion is weakly disadvantageous for any other type $\th'$. If type $\th'$ tries on test $\tau$ and his score is converted, he is weakly less likely to pass than if he tries on test $\psi$ directly.

In the language of statistical hypothesis testing, we can think of failing a test as rejecting the null hypothesis. Our definition requires that the conversion of test $\tau$ constitutes an hypothesis test of the null $\th$ against the alternative $\Th \setminus \{\th \}$ with significance $1 - \pi (\psi | \th)$ that is uniformly more powerful than test $\psi$. The requirement that $k_1 \geq k_0$ preserves incentives, which are not relevant in the statistical framework.

\begin{thm}[Test replacement] \label{res:replacement} Fix a type $\th$ and tests $\tau$ and $\psi$ such that $\tau \succeq_\th \psi$. If a social choice function is canonically implemented by a mechanism $(t,g)$ in which $t(\th) = \psi$, then it is also canonically implemented by some mechanism $(t', g')$ in which $t'(\th) = \tau$. 
\end{thm}

Here is a sketch of the proof. Start with a canonical implementation in which type $\th$ is given test $\psi$. Adjust the mechanism after the report $\th$ as follows. The principal gives the agent test $\tau$ and then converts the agent's score $s_\tau$ into a new score $s_\psi$ using the transition probabilities $k_1$ and $k_0$. Then the principal makes the decision that she would have made in the old mechanism after score $s_\psi$ on test $\psi$ (following report $\th$). 

This new mechanism implements the same social choice function. Suppose that type $\th$ reports truthfully and tries on test $\tau$. By \eqref{it:theta_equality}, he will get the same decision as in the equilibrium of the original mechanism. Suppose another type $\th'$ reports type $\th$ and tries on test $\tau$. By \eqref{it:theta_prime_inequality}, he will get a decision that he could have gotten in the original mechanism by reporting type $\th$ and then trying on test $\psi$ with some probability. The inequality $k_1 \geq k_0$ ensures that intentionally failing test $\tau$ also yields a decision that was achievable in the original mechanism. 

For each fixed type $\th$, the $\th$-discernment order $\succeq_\th$ is neither stronger nor weaker than \citeapos{Blackwell1953} order. Blackwell's order takes the same form as \cref{def:discernment_order} except (a) the inequality $k_1 \geq k_0$ is dropped, and (b) the inequality in \eqref{it:theta_prime_inequality} is strengthened to equality.  Blackwell's order is not suited to our setting because it does not consider the agent's incentives to intentionally fail a test. Indeed, Blackwell's order is invariant to relabeling the realizations ``pass'' and ``fail.'' Our $\th$-discernment order is not. 
 
Like Blackwell's order, each $\th$-discernment order $\succeq_\th$ is reflexive and transitive but not generally anti-symmetric. Tests $\tau_1$ and $\tau_2$ are \emph{$\th$-equivalent}, denoted $\tau_1 \sim_{\th} \tau_2$, if $\tau_1 \succeq_\th \tau_2$ and $\tau_1 \preceq_\th \tau_2$.  If two tests have the same passage rates, then they are clearly $\th$-equivalent. We show that the converse holds except in the special case that neither test can screen any other type away from type $\th$. Formally, type $\th$ is \emph{minimal} on test $\tau$ if $\pi(\tau | \th) \leq \pi (\tau | \th')$ for all types $\th'$. 
 
\begin{prop}[$\th$-discernment equivalence] \label{res:test_uniqueness} Fix a type $\th$. Tests $\tau_1$ and $\tau_2$ are $\th$-equivalent if and only if (a) $\pi (\tau_1 | \cdot) = \pi(\tau_2| \cdot)$ or (b) type $\th$ is minimal on $\tau_1$ and on $\tau_2$. 
\end{prop}

\subsection{Implementation with most-discerning testing} \label{sec:most_discerning}

\cref{res:replacement} is particularly useful if, for a given type $\th$, there is a single test that can replace every other test. 

\begin{defn}[Most-discerning] A test $\tau$ is \emph{most $\th$-discerning} if  $\tau \succeq_\th \psi$ for every $\psi$ in $T$. A function $t \colon \Th \to T$ is \emph{most-discerning} if for each type $\th$ the test $t(\th)$ is most $\th$-discerning. 
\end{defn}

Whether a test is most $\th$-discerning depends on the other tests in $T$. The only test that is more $\th$-discerning than \emph{every} test is the perfect test $\hat{\tau}_\th$ that exactly identifies whether the agent's type is $\th$, i.e., $\pi ( \hat{\tau}_\th | \th') = [\th' = \th]$, where $[ \cdot]$ is the indicator function for the predicate it encloses.

To state the main result, we define a \emph{decision environment} to consist of a decision set $X$ and a utility function $u \colon X \times \Th \to \R$ for the agent. 

\begin{thm} [Most-discerning implementation] \label{res:implementation} 
Fix a type space $\Th$ and a testing technology $(T, \pi)$. For a testing function $\hat{t} \colon \Th \to T$, the following are equivalent.
\begin{enumerate}
    \item \label{it:stat} $\hat{t}$ is most-discerning.
    \item \label{it:decision} In every decision environment $(X,u)$, every implementable social choice function can be canonically implemented with testing rule $\hat{t}$.
\end{enumerate}
\end{thm}

The forward implication from condition \ref{it:stat} to condition \ref{it:decision} says that a most-discerning
testing function suffices for all implementation problems. In the proof, for each type $\th$, we apply the procedure from \cref{res:replacement} to replace any test given to $\th$ with the test $\hat{t}(\th)$. 

The backward implication from condition \ref{it:decision} to condition \ref{it:stat} confirms that the most-discerning property is the right one. If $\tau \not\succeq_\th \psi$, then replacing test $\psi$ with test $\tau$ for type $\th$ introduces a new deviation outcome for some type. The proof constructs a decision environment in which this deviation outcome is profitable.

Even if the testing technology does not admit a most-discerning testing function, we can still use the replacement theorem (\cref{res:replacement}) 
to reduce the class of tests that need to be considered. Suppose there is a set $\hat{T}(\th)$ of tests with the following property: for every test $\psi$ there is some test $\tau$ in $\hat{T}(\th)$ such that $\tau \succeq_\th \psi$. Then there is no loss in assuming that the principal gives type $\th$ only
tests in $\hat{T}(\th)$, though the principal may randomize over tests in $\hat{T}(\th)$. See \cref{sec:most-discerning_correspondences} for a formal statement.

\begin{rem}[Discernment orders under alternative specifications] \label{rem:no_skipping}
Under the two alternative testing specifications described in \cref{sec:trying}---\emph{observable skipping} and \emph{exogenous scores}---the appropriate analogue of the $\th$-discernment order $\succeq_\th$ is Blackwell's order, for each $\th$ in $\Th$. With this redefinition of the discernment orders, it can be shown that \cref{res:replacement} and \cref{res:implementation} go through under each alternative specification.
\end{rem}

\subsection{Sufficient conditions for discernment orders} \label{sec:characterizing_discernment}

Checking whether one test $\tau$ is more-$\th$ discerning than another test $\psi$ amounts to verifying the feasibility of the system of linear inequalities in  \cref{def:discernment_order}. Here, we give a sufficient condition for $\th$-discernment in terms of relative performance. 

\begin{prop}[Relative performance] \label{res:relative_performance_tests}
Fix a type $\th$ and tests $\tau$ and $\psi$.\footnote{The two (non-exclusive) cases exclude the following two edge cases. If $\pi ( \tau| \th) > \pi(\psi | \th) = 0$, then $\tau \succeq_\th \psi$. If $\pi(\tau |\th) < \pi (\psi| \th) = 1$, then $\tau \not\succeq_\th \psi$, provided that $\pi (\psi| \cdot)$ is nonconstant.}
\begin{enumerate}
    \item \label{it:tau_easy} Suppose $\pi (\tau | \th) \geq \pi (\psi | \th) > 0$. Test $\tau$ is more $\th$-discerning than test $\psi$ if
    \begin{equation} \label{eq:char_tau_easy}
         \frac{\pi (\tau | \th')}{\pi (\tau | \th)} \leq \frac{ \pi (\psi | \th')}{\pi (\psi | \th)}, \quad \text{for all}~\th' \in \Th. 
    \end{equation}
    \item \label{it:tau_hard} Suppose $\pi (\tau | \th) \leq \pi (\psi | \th) < 1$. Test $\tau$ is more $\th$-discerning than test $\psi$ if
      \begin{equation} \label{eq:char_tau_hard}
         \frac{1 - \pi (\tau | \th')}{1 - \pi (\tau | \th)} \geq \frac{ 1 - \pi (\psi | \th')}{1 - \pi (\psi | \th)}, \quad \text{for all}~ \th' \in \Th.
    \end{equation}
\end{enumerate}
\end{prop}

In the first case, where type $\th$ is more likely to pass test $\tau$ than test $\psi$, test $\tau$ is more $\th$-discerning than test $\psi$ if for each type $\th'$ the relative \emph{passage} rate of type $\th'$ compared with type $\th$ is \emph{lower} on test $\tau$ than on test $\psi$. In the second case, where type $\th$ is more likely to fail test $\tau$ than test $\psi$, test $\tau$ is more $\th$-discerning than test $\psi$ if for each type $\th'$ the relative \emph{failure} rate of type $\th'$ compared with type $\th$ is \emph{higher} on test $\tau$ than on test $\psi$. 

\begin{rem}[Sufficient condition for most $\th$-discerning] \label{rem:another_sufficient} In view of \cref{res:relative_performance_tests}, a simple sufficient condition for test $\tau$ to be most $\th$-discerning is that $\tau$ maximizes $\pi (\cdot | \th)$ and $\tau$ minimizes $\pi ( \cdot | \th')$ for each type $\th'$ with $\th' \neq \th$. That is, among all tests in $T$, test $\tau$ is one that type $\th$ is most likely to pass but every other type is most likely to fail. 
\end{rem}

\section{Testing in reduced form} \label{sec:reduced_form} \label{sec:authentication_rate}

If the testing technology admits a most-discerning testing function, then the principal's design problem can be represented as a tractable optimization problem involving a single authentication rate $\a$. In this section, we analyze this reduction. 

Suppose that the testing technology $(T, \pi)$ admits a most-discerning testing function $\hat{t} \colon \Th \to T$. By \cref{res:implementation}, there is no loss of generality in restricting the principal to using  $\hat{t}$ as the testing rule. With this testing rule, the principal selects two decisions for each report $\th'$---the decision, $g_1(\th')$, if the agent passes test $\hat{t}(\th')$ and the decision, $g_0(\th')$, if the agent fails test $\hat{t}(\th')$. Suppose type $\th$ reports type $\th'$ and then tries on test $\hat{t}(\th')$.  With probability $\pi (\hat{t}(\th')| \th)$, he passes and gets $g_1(\th')$. With probability $1 - \pi(\hat{t}(\th')| \th)$, he fails and gets $g_0(\th')$. Define the induced authentication rate $\a$ by 
\begin{equation} \label{eq:induced_authentication}
\a ( \th' | \th) = \pi ( \hat{t}(\th') | \th), \quad \text{for all}~\th, \th' \in \Th.
\end{equation}
For any reduced outcome rule $g = (g_0, g_1) \colon \Th \to \D(X) \times \D (X)$, define the agent's associated utilities by
\[
    u(\th'|\th) = \a( \th' | \th) u(g_1(\th'), \th) + (1 - \a (\th'| \th)) u (g_0(\th'), \th), \quad \text{for all}~\th, \th' \in \Th.
\]

The principal's problem is to choose a reduced outcome rule $g$ to solve
\begin{equation} \label{eq:authentication_program}
\begin{aligned}
    &\text{maximize} &&\E[ \a(\th | \th) v( g_1 (\th) , \th) + (1 - \a(\th| \th)) v( g_0(\th), \th)] \\
    &\text{subject to} &&u ( \th | \th) \geq u (\th' | \th) \vee u(g_0(\th'), \th), \quad \text{for all}~ \th, \th' \in \Th.
\end{aligned}
\end{equation}
The constraints capture truthtelling and obedience. They require that for each type $\th$, reporting $\th$ and trying on test $\hat{t}(\th)$ is weakly preferred to reporting any type $\th'$ and either trying on test $\hat{t}(\th')$ or intentionally failing it. In particular, with $\th' = \th$, the constraint ensures that type $\th$ weakly prefers to try on test $\hat{t}(\th)$ rather than intentionally failing it. In contrast to models of exogenous lying costs,%
\footnote{In models of lying costs, reports have literal meanings. The agent pays a cost $c(\th' |\th)$ if he reports $\th'$ when his true type is $\th$. See, for example, \cite{LackerWeinberg1989},
\cite{MaggiRodriguez-Clare1995},
\cite{CrockerMorgan1998},
\cite{KartikOttavianiSquintani2007},
\cite{Kartik2009}, 
and
\cite{DeneckereSeverinov2017wp}.
Within mechanism design, \cite{KephartConitzer2016} show that if the lying cost function satisfies the triangle inequality, then there is no loss in restricting to truthful equilibria.}
here the effective cost of misreporting is determined jointly by the authentication rate and the principal's choice of decisions when the agent is not authenticated.

The approach described above motivates the following definition. 
 
\begin{defn}[Most-discerning authentication rate] An authentication rate $\a$ is \emph{most-discerning} if there exists a testing technology $(T, \pi)$ with a most-discerning testing function $\hat{t}$ such that
\begin{equation} \label{eq:induced}
    \a ( \th' | \th) = \pi ( \hat{t} (\th') | \th), \quad \text{for all}~\th, \th' \in \Th. 
\end{equation}
\end{defn}

If $\a$ is most-discerning, then we can directly study the program \eqref{eq:authentication_program}, with the assurance that it represents the designer's full problem for some testing technology. 

Now suppose that an \emph{arbitrary} authentication rate $\a$ is specified directly; for examples of this approach, see \cite{CaragiannisEtal2012} and \cite{FerraioliVentre2018}. As long as communication is unrestricted, we believe the natural interpretation of a primitive authentication rate $\a$ is that for each report $\th'$ there is an associated test, which each type $\th$ can pass with probability $\a ( \th' | \th)$. Formally, the principal has available the testing technology $(T^\a, \pi^\alpha)$, defined by
\begin{equation} \label{eq:induced_testing}
    T^\a = \{ \tau^\a_{\th'}: \th' \in \Th\}, 
    \qquad
    \pi^\a (\tau^\a_{\th'} | \th) = \a ( \th'| \th).
\end{equation}
If this construction is applied to \cref{sec:cautionary-example}, then the testing function 
$\th \mapsto \tau_\th^\alpha$ is not most-discerning.

\begin{rem}[Most-discerning] \label{rem:authentication_linear_system}
It is easily verified that an authentication rate $\a$ is most-discerning if and only if, under the associated testing technology $(T^\alpha, \pi^\alpha)$, the testing function $\th \mapsto \tau_\th^\a$ is most-discerning.\footnote{If there exists a testing technology $(T, \pi)$ with a most-discerning testing function $\hat{t}$ that satisfies \eqref{eq:induced}, then for all distinct types $\th$ and $\th'$, we have $\tau_\th^\a \sim \hat{t}(\th) \succeq_\th \hat{t}(\th') \sim_\th \tau_{\th'}^\a$.} Therefore, an authentication rate $\a$ is most-discerning if and only if for all distinct types $\th$ and $\th'$, we have $\tau_\th^\a \succeq_\th \tau_{\th'}^{\alpha}$, i.e., there exist $k_0  = k_0 (\th ,\th')$ and $k_1 = k_1 ( \th, \th')$ with $0 \leq k_0 \leq k_1 \leq 1$ satisfying
\begin{equation} \label{eq:alpha_system_main_text}
\begin{aligned}
    \a ( \th | \th) k_1 + (1 - \a ( \th | \th)) k_0 &= \a ( \th' | \th), \\
    \a ( \th | \th'') k_1 + (1 - \a ( \th | \th'')) k_0 &\leq \a ( \th' | \th''), \quad \text{for all}~\th'' \in \Th \setminus \{\th\}.
\end{aligned}
\end{equation}
\end{rem}

By \cref{rem:authentication_linear_system},  checking whether an authentication rate $\a$ is most-discerning amounts to verifying whether a particular system of linear inequalities is feasible. We now give a simpler characterization of whether an authentication rate $\a$ is most-discerning, under one additional assumption on $\a$.

\begin{prop}[Most-discerning characterization] \label{res:most_discerning_authentication}
Let $\a$ be an authentication rate satisfying $\a (\th |\th) \geq \max \{ \a (\th' | \th), \a (\th | \th') \}$ for all $\th, \th' \in \Th$. Then $\a$ is most-discerning if and only if
\begin{equation} \label{eq:alpha_NRC}
    \a (\th_3 | \th_2) \a( \th_2 | \th_1) \leq \a (\th_3 | \th_1) \a(\th_2 | \th_2), \quad \text{for all}~\th_1, \th_2, \th_3 \in \Th.
\end{equation}
\end{prop}

If $\a (\th | \th) =1$ for all types $\th$, then it follows from \cref{res:most_discerning_authentication} that $\a$ is most-discerning if and only if $\a$ is \emph{supermultiplicative}: for all types $\th_1, \th_2, \th_3$, it is more likely that $\th_1$ is directly authenticated as type $\th_3$ than that type $\th_1$ is authenticated as type $\th_2$, and type $\th_2$ is (independently) authenticated as type $\th_3$.

To be sure, the principal's design problem cannot always be reduced to an optimization problem of the form \eqref{eq:authentication_program}.
If the testing technology does not admit a most-discerning testing function, then the full protocol from \cref{fig:timing} must be considered.\footnote{In many cases,  the class of testing rules can still be reduced; see \cref{sec:most-discerning_correspondences}.} Similarly, if an authentication rate $\a$ is not most-discerning, then a solution of \eqref{eq:authentication_program} need not be optimal among all dynamic mechanisms that use the associated testing technology $(T^\a, \pi^\a)$. Still, there are many settings that can be reduced to the program in  \eqref{eq:authentication_program}. We conclude this section with a few examples.

\begin{exmp}[Evidence verification with error]
Suppose that each type $\th$ has a distinct piece of evidence $e_\th$. The agent chooses whether to present the evidence he possesses. The principal has a system to check whether the presented evidence matches what was requested. A mismatch is detected with probability $1 - \e$, where $0 < \e < 1$.  

For each type $\th'$, let $\tau_{\th'}$ denote the test that requests evidence $e_{\th'}$. The agent passes this test if he presents a piece of evidence, and the system does not detect a mismatch between the presented evidence and $e_{\th'}$. On this test, type $\th$ can ``try'' by presenting evidence $e_\th$ or ``intentionally fail'' by presenting no evidence. The associated passage rate is given by 
\[
    \pi ( \tau_{\th'} | \th) =     
    \begin{cases}
        1 &\text{if}~\th = \th', \\
        \e &\text{if}~\th \neq \th'.
    \end{cases}
\]
The map $\th \mapsto \tau_\th$ is most-discerning; to see this, check the sufficient condition in \cref{rem:another_sufficient}. The authentication rate representing this technology is analyzed in \cite{dziuda2018} and \cite{balbuzanov2019}.
\end{exmp}

\begin{exmp}[Semi-metric authentication rate] Let $d$ be a semi-metric on the type space $\Th$.\footnote{Unlike a metric, a semi-metric does not require that $d(\th, \th') \neq 0$ for $\th \neq \th'$.} Consider the authentication rate $\a$ defined by
\[
    \a ( \th' | \th) = \exp \{- d ( \th, \th')\}.
\]
The interpretation is that types that are closer in the semi-metric $d$ are more similar and hence are more difficult to distinguish.\footnote{As a special case, suppose $\Th = \R^k$ and $d$ is induced by a norm. The resulting class of functions $\a$ is axiomatized in \cite{BillotEtal2008}. They interpret these functions as measures of similarity in their model of belief-formation through similarity-weighted averaging.} Using \cref{res:most_discerning_authentication}, it is easy to check that $\a$ is most-discerning.
\end{exmp}


\begin{exmp}[Separate verifiable and payoff components] \label{ex:two_components} Let $\Th = \Th^0 \times \Th^1$, with a generic type denoted by $\th = ( \th^0, \th^1)$. Let $\a^0$ be a most-discerning authentication rate on $\Th^0$. We obtain a most-discerning authentication rate $\a$ on $\Th$ by defining 
\[
    \a  ( \hat{\th}^0, \hat{\th}^1 |  \th^0, \th^1) = \a^0 ( \hat{\th}^0 | \th^0),
\]
for all $\th^0, \hat{\th}^0 \in \Th^0$ and all $\th^1, \hat{\th}^1 \in \Th^1$. Think of $\th^0$ as a verifiable attribute, and $\th^1$ as an unverifiable payoff type. \cite{Reuter2023} considers this structure in a model of partial verification.  The prior distribution over $\Th^0 \times \Th^1$ determines how informative the agent's verifiable attribute is about his payoff type. For example, the verifiable attribute could indicate whether the agent is a student, and the valuation distribution among students may be different from the valuation distribution among non-students. 
\end{exmp}

\begin{exmp}[Partial verification] \label{ex:partial_verification}
Suppose that the authentication rate $\a$ is $\{0,1\}$-valued. For each type $\th$, let $M(\th) = \{ \th' \in \Th : \a (\th' | \th) = 1\}$. Hence, type $\th$ can be authenticated as any type in $M(\th)$. Following \cite{GreenLaffont1986}, suppose that each type can be authenticated as himself, i.e, $\a ( \th | \th) = 1$ for all $\th$. In terms of $M$, \eqref{eq:alpha_NRC} becomes
\[
    \th_3 \in M(\th_2) \quad \& \quad \th_2 \in M(\th_1) \implies \th_3 \in M(\th_1).
\]
This is exactly \citeapos{GreenLaffont1986} nested range condition. Under this condition, they argue that the revelation principle is valid. 
\end{exmp}

\section{Applications: Profit-maximization with verification}
 \label{sec:applications}

As an application of the reduced-form representation of the principal's design problem, we solve for profit-maximizing mechanisms with verification in a few classical settings. 

\subsection{Setting}

\paragraph{Authentication rate}
The type space is an interval $\Th = [\ubar{\th}, \bar{\th}]$, where $\bar{\th} > \ubar{\th} \geq 0$. We represent the verification technology by a most-discerning authentication rate $\a$. Assume that $\a$ takes the exponential form
\begin{equation} \label{eq:exponential_form}
    \a ( \th' | \th) = \exp \Paren{ - \Abs{ \int_{\th'}^{\th} \l(\xi) \de \xi} }, \quad \text{for all}~\th,\th' \in \Th,
\end{equation}
for some continuous function $\l \colon [\ubar{\th}, \bar{\th}] \to \R_+$. It is easily verified that this function $\a$ is most-discerning. The exponential authentication rate allows for a cleaner characterization of optimal mechanisms. With other most-discerning authentication rates, additional regularity conditions are needed to ensure that global deviations are not profitable; see \cref{sec:beyond_exponential}.

The parameter $\l(\th)$ quantifies the local precision of the verification technology near type $\th$. The function $\a(\th | \cdot)$ has a kink at type $\th$ if and only if $\l(\th) > 0$. \cref{fig:authentication_rates} plots the authentication rate when $\l(\th) = 1$ for all $\th$. The plot shows the authentication probability, as a function of the agent's true type, for two fixed reports $\th'$ and $\th''$.

\begin{figure}
\centering
\begin{tikzpicture}
	\begin{axis}[
		axis lines = left,
		xmin = 0,
		xmax = 1.1,
		ymin = 0,
		ymax = 1.1,
		xtick = {0,0.25, 0.75, 1},
		xticklabels = {0,$\th'$, $\th''$, 1},
		ytick = {0,1},
		yticklabels = {0,1},
		xlabel = {$\th$},
		xlabel style={at=(current axis.right of origin), anchor=west},
		clip = false
	]
	
	\addplot [domain = 0:0.25] {e^(x - 0.25)};
	\addplot [domain = 0.25:1] {e^(0.25 - x)}
	node [pos = 1, anchor = west] {$\alpha(\th' | \th)$};
	\addplot [dotted] coordinates { (0.25,0) (0.25,1)};	
	
	\addplot [domain = 0:0.75] {e^(x - 0.75)};
	\addplot [domain = 0.75:1] {e^(0.75 - x)}
	node [pos = 1, anchor = west] {$\a(\th'' | \th)$};
	\addplot [dotted] coordinates { (0.75,0) (0.75,1)};	
	\end{axis}
\end{tikzpicture}	
\caption{Exponential authentication rate}
\label{fig:authentication_rates}
\end{figure}

\paragraph{Quasilinear environment}

The agent's type $\th \in \Th = [\ubar{\th}, \bar{\th}]$ is drawn from a distribution function $F$ with strictly positive density $f$. The principal allocates a quantity $q \in  Q \subset \R_+$ and receives a transfer $t \in \R$.%
\footnote{The pair $(q,t)$ corresponds to the decision $x$ in the general model. Throughout \cref{sec:applications}, $t$ always denotes transfers (and we make no direct reference to tests).}
The set $Q$ will be either $[0,1]$ or $\R_+$, depending on the application. Utilities for the agent and the principal are given by
\[
    u(q,t,\th) = \th q - t
    \quad
    \text{and}
    \quad
    v(q,t) = t - c(q),
\]
for some weakly convex cost function $c \colon Q \to \R_+$. For concreteness, we interpret the principal as the seller of a good and the agent as a potential buyer.\footnote{An alternative interpretation of this setting is that the principal is the procurer of a good who can imperfectly verify the agent's production costs.} The agent's type represents his market segment, which the principal can imperfectly verify. To avoid the difficulties of mechanism design with multi-dimensional types, we make the stylized assumption that the agent's market segment pins down his valuation. With the specified authentication rate, it is more difficult for the seller to distinguish buyers who are in market segments with closer valuations. 

The agent is free to walk away at any time, so we impose an ex-post participation constraint.%
\footnote{Formally, after the agent observes the test result, he has the right to walk away,  free and clear, with no payment obligation. This assumption rules out upfront payments like those used in \cite{BorderSobel1987}.}
If the principal could impose arbitrarily severe punishments for failed authentication, then probabilistic verification would be essentially as effective as perfect verification; see \cite{CaragiannisEtal2012}.

Since $\a(\th |\th) = 1$ for all $\th$, the agent is always authenticated if he is truthful. Therefore, failed authentication is off path. Given the ex-post participation constraint, we may assume without loss that if the agent fails to be authenticated, then the principal excludes him---the agent pays nothing and does not get the good. Formally, we set $g_0( \th) = (0,0)$ for all $\th$, and we optimize over the decision rule $g_1$. Denote the quantity and transfer components of $g_1$ by $q$ and $t$.

The principal selects a quantity function $q \colon \Th \to Q$ and a transfer function $t \colon \Th \to \R$ to solve
\begin{equation} \label{eq:optimization}
\begin{aligned}
	&\operatorname{maximize}  && \int_{\ubar{\th}}^{\bar{\th}} [ t(\th) - c(q(\th))]  f(\th) \de \th  \\
	&\text{subject to} && \th q(\th) - t(\th) \geq \a(\th' | \th) [ \th q(\th') - t(\th')], \quad \text{for all}~\th, \th' \in \Th \\
	&&&\th q(\th) - t(\th) \geq 0, \quad \text{for all}~\th \in \Th.
\end{aligned}
\end{equation}
Here, the constraints from \eqref{eq:authentication_program}
take a simple form because $u(g_0(\th'), \th) = 0$ for all $\th$ and $\th'$. These constraints guarantee ex-post participation.\footnote{Since $g_0(\th') = (0,0)$ for all $\th'$, the agent gets his outside option whenever he fails to be authenticated. So in this case, the constraints in \eqref{eq:authentication_program} imply the ex-post participation constraints.}

\subsubsection{Virtual value}

We derive a new expression for the virtual value in this quasilinear setting with verification. In the classical setting without verification, the envelope theorem pins down (almost everywhere) the derivative of the agent's indirect utility function $U$  in terms of the allocation rule: $U'(\th) = q(\th)$. Hence, 
\begin{equation} \label{eq:U_noV}
    U(\th) = U(\ubar{\th}) + \int_{\ubar{\th}}^{\th} q(\xi) \de \xi \geq \int_{\ubar{\th}}^{\th} q(\xi) \de \xi.
\end{equation}
With verification, the derivative of the agent's indirect utility function $U$ is no longer pinned down by the quantity function because of the kink in $\a(\th |\cdot)$. Instead, the envelope formula gives the differential inequality\footnote{See \cite{CarbajalEly13} for a general characterization of indirect utility functions, when the agent's primitive utility function is kinked. \cite{CarbajalEly16} apply this characterization in a model of reference-dependent utility.}
\begin{equation} \label{eq:diff_ineq}
    q(\th) - \l(\th) U(\th) \leq U'(\th) \leq q(\th) +\l(\th) U(\th).
\end{equation}
This differential inequality depends only on the \emph{local} behavior of $\a$ around the diagonal, which is captured by the function $\l$. Indeed, the left and right derivatives of the function $\a ( \th| \cdot)$, evaluated at $\th$, equal $\l(\th)$ and $-\l(\th)$, respectively. The greater the local verification precision $\l(\th)$, the more permissive is the inequality in \eqref{eq:diff_ineq}.

The lower bound in \eqref{eq:diff_ineq}  can be shown to imply the bound
\begin{equation} \label{eq:U_V}
    U(\th) \geq \int_{\ubar{\th}}^{\th} e^{ -\int_{\xi}^{\th} \l(z) \de z } q(\xi) \de \xi.
\end{equation}
The right side of \eqref{eq:U_V} solves the differential equation $U'(\th) = q(\th) - \l (\th) U(\th)$. Since $\a$ takes the exponential form in \eqref{eq:exponential_form}, the integrand in \eqref{eq:U_V} reduces to $\a( \xi | \th) q(\xi)$. We will use this simpler expression below, but remember that the solution is pinned down by the envelope formula, not by global deviations.

It is optimal to choose $U$ so that \eqref{eq:U_V} holds with equality. After substituting this choice of $U$ into the objective and changing the order of integration, the principal's objective can be expressed as a linear functional in $q$. The coefficient on $q(\th)$ is the virtual value of type $\th$:
\[
    \varphi(\th) = \th - \frac{1}{f(\th)} \int_{\th}^{\bar{\th}} \a ( \th| \xi) f(\xi) \de \xi.
\]
Myerson's virtual value is derived similarly in the no-verification problem, using \eqref{eq:U_noV} in place of \eqref{eq:U_V}. Myerson's virtual value can be expressed symmetrically as
\[
    \varphi^M(\th) = \th - \frac{1}{f(\th)} \int_{\th}^{\bar{\th}} f(\xi) \de \xi.
\]

The virtual value of type $\th$ captures the marginal revenue from allocating to type $\th$. It has two parts. First, the principal can extract the consumption utility $\th$ from type $\th$. Second, the allocation pushes up the indirect utility of each type $\xi$ with $\xi > \th$. This marginal effect on type $\xi$, which equals $1$ (without verification)  and $\a(\th |\xi)$ (with verification), is then integrated against the relative density $f(\xi)/ f(\th)$. Comparing the virtual values,  we immediately see that
\[
	\varphi^{M} (\th) \leq \varphi(\th) \leq \th.
\]
The virtual value $\varphi(\th)$ tends toward these bounds in the limiting cases. As $\l$ converges to $0$ pointwise, $\varphi(\th)$ converges to $\varphi^M(\th)$ for each type $\th$. Conversely, as $\l$ converges to $\infty$ pointwise, $\varphi (\th)$ converges to $\th$ for each type $\th$. 

Below, we will characterize optimal mechanisms under the assumption that the virtual value $\varphi$ is increasing. This is a joint assumption on the type distribution and the authentication rate. If $\l ( \th) = \l$ for all $\th$, then the virtual value has a simple expression for some standard distributions. In particular, for both uniform and exponential distributions, the virtual value is strictly increasing.

\subsection{Optimal mechanisms}

We find the optimal mechanism in two classical problems. 

\subsubsection{Nonlinear pricing}

For nonlinear pricing \citep{MussaRosen1978}, 
the quantity space is $Q = \R_+$. Assume that the principal's cost function $c$ satisfies the standard assumptions: $c'(0)=0$, the derivative $c'$ is strictly increasing, and $\lim_{q \to \infty} c'(q) > \bar{\th}$. Say that the optimal mechanism is \emph{essentially unique} if all optimal mechanisms agree at almost every type. 

\begin{prop}[Optimal nonlinear pricing] \label{res:nonlinear_optimum}
Assume that the virtual value $\varphi$ is weakly increasing. The optimal quantity function $q^\ast$ and transfer function $t^\ast$ are essentially unique and given by
\begin{equation*} \label{eq:nonlinear_optimum}
	c'(q^\ast(\th)) = \varphi(\th)_+,
	\qquad
	t^\ast(\th) = \th q^\ast(\th) - \int_{\ubar{\th}}^{\th} \a(\xi | \th) q^\ast(\xi) \de \xi.
\end{equation*}
\end{prop}

The optimal allocation rule has the same form as in the classical case, except the new virtual value appears in place of the classical virtual value. Transfers are determined by the indirect utility function $U$, which is given by the minimal solution of \eqref{eq:U_V}. 

Each type $\th$ receives the quantity that is efficient for type $\varphi(\th)_+$. Therefore, quantity is distorted below the efficient level for every type except $\bar{\th}$. As the verification precision $\l$ increases pointwise, 
downward distortion is attenuated. In the limit of perfect verification, the good is allocated efficiently and the principal extracts the full surplus. 

\subsubsection{Selling a single indivisible good}

For a single indivisible good \citep{RileyZeckhauser1983}, the quantity space is $Q = [0,1]$. Here, quantity is interpreted as the probability of allocating the good. Hence, $c(q) = c q$, where $c$ is the cost of producing a single good. Assume $0 \leq c < \bar{\th}$. 

Without verification, the profit-maximizing mechanism is a posted price. With verification, the seller charges different prices to consumers in different market segments. 

\begin{prop}[Optimal sale of a single good] \label{res:posted_price}  Assume that the virtual value $\varphi$ is strictly increasing. The optimal quantity function $q^\ast$ and transfer function $t^\ast$ are essentially unique and given as follows. Let $\th^\ast = \varphi^{-1}(c)$. If $\th < \th^\ast$, then $q^\ast(\th) = t^\ast(\th) = 0$. If $\th \geq \th^\ast$, then $q^\ast(\th) = 1$ and 
\[
    t^\ast (\th) = \th^\ast + \int_{\th^\ast}^{\th} (1 - \a ( \xi | \th)) \de \xi.
\]
\end{prop}

As in the classical solution, the allocation probability takes values $0$ and $1$ only---there is no randomization.\footnote{\cite{SherVohra2015} study this selling problem with deterministic evidence, assuming the type space is finite. The optimal mechanism may involve lotteries. They give a condition on the evidence structure under which the optimal mechanism is deterministic.} There is a cutoff type $\th^\ast$ who receives the good and pays his valuation. Each type below the cutoff is excluded. Each type above the cutoff receives the good and pays a price that is less than his valuation. The price is no longer uniform. As long as $\l$ is strictly positive, the price is strictly increasing in the agent's report. Nevertheless, types above the cutoff cannot profit by misreporting downward---the benefit of a lower price is outweighed by the risk of failing to be authenticated and getting nothing. As verification becomes more precise, the price becomes more sensitive to the agent's type, and more types receive the good. 

\begin{rem}[Auctions] Our model can be extended to allow for multiple agents.\footnote{In this extension, we assume that the principal tests the agents simultaneously. In particular, the test given to one agent cannot depend on another agent's test score.} In the revenue-maximizing auction, the allocation rule takes the familiar form from \citet{Myerson1981}, with our generalized virtual value in place of the classical virtual value. In the asymmetric case, the allocation rule favors a bidder if his valuation distribution is lower or if his valuation can be verified more precisely. 
\end{rem}

\section{Beyond pass--fail tests} \label{sec:nonbinary}

The main model considers pass--fail tests. In this section, we consider tests that generate scores in a finite score set $S$. The agent's type-dependent performance on each test is represented by a map $\pi \colon T \times \Th \to \D (S)$,
which specifies for each type $\th$ and test $\tau$ a distribution $\pi_{\tau |\th}$ over $S$. 

To generalize the agent's choice of whether to try on a test, we take as primitive a partial order $\succeq$ on $S$. The interpretation is that the agent can shift probability from score $s$ to score $s'$ if and only if $s \succeq s'$. As before, the agent's choice is costless, and the principal observes only the final score. The main model of pass--fail testing corresponds to $S = \{0,1\}$ with the usual order $\succeq$; by mixing, type $\th$ can choose to pass test $\tau$ with any probability below $\pi (\tau | \th)$. In the general model, type $\th$ can achieve on test $\tau$ any score distribution $p$ in $\D(S)$ satisfying $\pi_{\tau | \th} \succeq_{\ST} p$, where $\succeq_{\ST}$ is the stochastic order between probability measures on the partially ordered space $(S, \succeq)$. That is, $\mu \succeq_{\ST} \nu$ if and only if  $\mu (U) \geq \nu(U)$ for every upper set $U$;\footnote{An \emph{upper set} is a set with the property that if $s$ is in $U$ and $s' \succeq s$, then $s'$ is also in $U$.} see \cite{KamaeEtal1977}.

We define the $\th$-discernment orders in this more general setting. A function $k \colon S \to \D(S)$ is increasing if $k(s) \succeq_{\ST} k(s')$ whenever $s \succeq s'$. We interpret $k$ as a Markov transition, and we use the following notation from Markov chains. Given $\mu$ in $\D(S)$ and $k \colon S \to \D(S)$, the measure $\mu k$ on $\D(S)$ is defined by $(\mu k) (A) = \sum_{s} \mu(s) k(A |s)$, for $A \subset S$. 

\begin{defn}[$\th$-discernment for general tests] \label{def:discernment_order_general} 
Fix a type $\th$. Test $\tau$ is \emph{more $\th$-discerning} than test $\psi$, denoted $\tau \succeq_\th \psi$, if there exists an increasing function $k \colon S \to \D(S)$ such that
\begin{enumerate}[label = (\roman*)]
    \item \label{it:theta_equality_gen}  $\pi_{\tau|\th} k = \pi_{\psi|\th}$;
    \item \label{it:theta_prime_inequality_gen} $\pi_{\tau|\th'} k \preceq_{\ST} \pi_{\psi | \th'}$ for all types $\th'$ with $\th' \neq \th$. 
\end{enumerate}
\end{defn}

This order $\succeq_\th$ is reflexive and transitive; see \cref{sec:nonbinary_appendix} for a proof. We can define \emph{most $\th$-discerning tests} and \emph{most-discerning testing functions} with respect to this definition of $\succeq_\th$. With this generalized testing technology, the revelation principle (\cref{res:revelation_principle}), the replacement theorem (\cref{res:replacement}), and the forward implication in the main implementation theorem (\cref{res:implementation}) go through with similar proofs.\footnote{A most-discerning testing function $\hat{t}$ induces a generalized authentication rate $\a \colon \Th \times \Th \to \D(S)$ defined by $\a ( \th' | \th) = \pi_{\hat{t}(\th') | \th}$. We can set up an analogue of the program in \eqref{eq:authentication_program}, but the incentive constraints depend on the order $\succeq$ on $S$.}

\section{Related literature on verification} \label{sec:discussion}

Verification has been modeled in many ways, in both economics and computer science. Here, we focus on costless, imperfect verification.%
\footnote{In economics, ``verification'' traditionally means that the principal
can learn the agent's type perfectly by taking some action, e.g., paying a fee or allocating a good. This literature began with \cite{Towsend1979} who studies \emph{costly verification} in debt contracts. 
\cite{BenPorathDekelLipman2019} connect costly verification and evidence. When monetary transfers are infeasible, costly verification is often used as a substitute; see \cite{BenPorathDekelLipman2014}, \cite{MylovanovZapechelnyuk2017}, \cite{ErlansonKleiner2015}, \cite{HalacYaredWP}, and \cite{li2017}.}

\cite{GreenLaffont1986} introduce \emph{partial verification}.%
\footnote{A precursor of their work is \cite{Postlewaite1979}, which considers exchange mechanisms when endowments are hidden. Each agent can benefit by withholding (and consuming) part of his endowment.} They restrict their analysis to direct mechanisms. Verification is represented as a correspondence $M \colon \Th \twoheadrightarrow \Th$ satisfying $\th \in M(\th)$ for each type $\th$.  Each type $\th$ can ``report'' any type $\theta'$ in $M(\th)$. This correspondence $M$ can be represented within our model as a $\{0,1\}$-valued authentication rate. We reinterpret the apparent failure of the revelation principle in \citeapos{GreenLaffont1986} model as a consequence of taking as primitive an authentication rate that is not most-discerning; see \cref{ex:partial_verification} for the formal connection.

\cite{BullWatson2004,BullWatson2007}, \cite{DeneckereSeverinov2008}, and \cite{LipmanSeppi1995} study \emph{hard evidence}.%
\footnote{Evidence was introduced in games (without commitment) by \cite{Milgrom1981} and \cite{Grossman1981}; for more recent work on evidence games, see \cite{HartKremerPerry2017}, \cite{BenPorathLipman2017}, and \cite{KoesslerPerez-Richet2017wp}.}
They introduce an abstract evidence set $\mathcal E$ and an evidence correspondence $E \colon \Theta\twoheadrightarrow \mathcal E$. Type $\th$ possesses the evidence in $E(\th)$, and he can present one piece of evidence from $E(\th)$.  Presenting evidence is costless.\footnote{In \cite{KartikTercieux2012}, the agent can provide evidence at a cost, which depends on the state. The focus of their paper is full implementation.} \cite{BullWatson2007} show that this evidence model can be represented in a reduced form if the evidence environment is \emph{normal}, i.e., each type $\th$ has a piece of evidence $e(\th)$ in $E(\th)$ that is \emph{maximal} for type $\th$ in the following sense: every other type $\th'$ who has $e(\th)$ also has every other piece of evidence in $E(\th)$. This model of deterministic evidence can be represented within our model as follows. For each piece of evidence $e$ in $\EE$, define the test $\tau_e$ that requests evidence $e$. Type $\th$ can pass test $\tau_e$ if and only if $e$ is in $E(\th)$. Every type can intentionally fail any test by withholding his evidence.\footnote{Formally, in \cite{BullWatson2007}, the agent \emph{must} present a piece of evidence from $E(\th)$. Disclosing nothing can be represented in their framework as a distinguished piece of evidence that every type possesses. If there exists such ``minimal evidence'' \citep[p.~85]{BullWatson2007}, then their evidence model is equivalent to our testing representation.} A piece of evidence $e$ in $E(\th)$ is maximal for type $\th$ in the sense of \cite{BullWatson2007} if and only if test $\tau_e$ is most-$\th$ discerning in our sense.\footnote{\cite{Strausz2016wp} revisit the setting of \cite{BullWatson2007}. They consider both the standard specification (termed ``non-controllable evidence'') and an alternative specification (``controllable evidence'') in which contracts can be written demanding particular evidence provision. They give conditions under which controllability has no value for the principal.}

In computer science, \cite{CaragiannisEtal2012} and \cite{FerraioliVentre2018} consider a primitive authentication rate, and they restrict attention to truthful equilibria of direct mechanisms. Our paper shows that the restriction to direct, truthful mechanisms is without loss if $\a$ is most-discerning. \cite{CaragiannisEtal2012} allow the principal to use arbitrarily severe punishments to deter any report that is not authenticated with certainty. In our applications (\cref{sec:applications}), the agent can walk away at any time, so punishment is limited to the agent's outside option.

Closest to our paper is the independent paper of \cite{BenPorathDekelLipmanWP}. They consider an abstract evidence set $\mathcal E$. In their \emph{signal-choice model},\footnote{They also consider a more general \emph{evidence-acquisition model}, and they give conditions under which a given evidence-acquisition model can be represented as a signal-choice model.} the primitive is a correspondence $A \colon \Theta\twoheadrightarrow \D(\mathcal E)$. Type $\th$ can choose any distribution $a$ in $A(\th)$. Then evidence $e$ in $\EE$ is realized according to the distribution $a$. For each type $\th$, they define an associated informativeness order over $A(\th)$, which depends on the full correspondence $A$. In the spirit of our \cref{res:replacement}, they show that for implementation, each type's equilibrium choice of a less informative distribution can be replaced with a more informative distribution. It is possible to embed our testing protocol in their signal-choice model.\footnote{Given a nonbinary testing technology $(T,S,\pi)$, consider their signal-choice model with $\EE = T \times S$. For each type $\th$, let $A(\th)$ be the set of distributions $\d_{\tau} \otimes p$ for all $\tau \in T$ and $p \in \D(S)$ satisfying $\pi_{\tau | \th} \succeq_{\ST} p$. Under this embedding, $\tau$ is more $\th$-discerning than $\psi$ in our framework if and only if $\d_{\tau} \otimes \pi_{\tau | \th}$ is more informative (for type $\th$) than $\d_{\tau} \otimes \pi_{\psi | \th}$ in theirs.} Our papers have different aims. \cite{BenPorathDekelLipmanWP} study the relationship between different evidence protocols in the most general setting. We impose more structure in order to obtain a tractable verification framework that we can apply to classical mechanisms design problems. 

\section{Conclusion} \label{sec:conclusion}

We model probabilistic verification as a technology---a family of tests that are available to the principal. The principal chooses how to use this testing technology within an arbitrary dynamic mechanism. We characterize whether this complex problem can be reduced to a static problem with relaxed incentive constraints. Then we solve this reduced problem using the first-order approach in a few classical profit-maximization applications. We believe this first-order approach will be useful for solving other mechanism design problems with probabilistic verification. 

We have found the optimal mechanism for each fixed verification technology in a nonparametric family. We can therefore quantify the value of each technology to the principal. This is the first step towards analyzing a richer setting in which the principal chooses how much to invest in verification technologies. We leave this to future research.

\appendix

\newpage
\section{Proofs} \label{sec:main_proofs}

\subsection{Proof of \texorpdfstring{\cref{res:revelation_principle}}{Proposition \ref{res:revelation_principle}}}

Let $S = \{0,1\}$. Consider a mechanism $(M, M'; t, r', g)$ and a strategy $(r, a)$. For each fixed type $\th$, the sequence $(m, \tau, m', s', x)$ in $M \times T \times M' \times S \times X$ is realized according to the following procedure. (Below, the symbol $\sim$ denotes ``distributed according to.'')
\begin{itemize}
    \item Agent sends $m \sim r (\th)$.
    \item Principal selects $\tau \sim t(m)$.
    \item Principal sends $m' \sim r' ( m, \tau)$.
    \item Agent tries with probability $a(\th, m, \tau, m')$.
    \item Nature draws $s'$ according to $\pi(\tau | \th)$ and whether the agent tried.
    \item Principal selects $x \sim g(m, \tau, m', s')$.
\end{itemize}
This distribution of $(m, \tau, m', s', x)$ is replicated by the following canonical procedure:
\begin{itemize}
    \item Agent sends $\th'  = \th$.
    \item Principal privately draws $m \sim r (\th')$ and then selects $\tau \sim t(m)$.
    \item Agent tries.
    \item Nature draws $s$ according to $\pi(\tau | \th)$ and whether the agent tried.
    \item Principal privately draws $m' \sim r'( m, \tau)$; then privately draws $s' \in \{0,1\}$ so that $s' = 1$ with probability $a(\th', m, \tau, m') s$;  and finally selects $x \sim g( m, \tau, m', s')$.
\end{itemize}
We check that the outcome of any deviation by type $\th$ in the new mechanism can be replicated by a deviation in the old mechanism. It follows that such a deviation cannot be profitable. If type $\th$ (i) reports $\th' \sim \rho$ and (ii) tries with probability $\a(\th', \tau)$, this can be replicated in the old mechanism by (i) privately drawing $\th' \sim \rho$ and then sending $m \sim r(\th')$; and (ii) trying with probability $a ( \th', m, \tau, m') \a(\th', \tau)$.%
\footnote{This argument relies in two places on a form of randomization that our model does not technically allow. In the canonical mechanism, the principal remembers her privately drawn $m$ and uses it to select $x$.  In the replicating deviation in the original mechanism, the agent remembers his privately drawn $\th'$ before choosing whether to try. We can replace this memory with fresh draws from the correct conditional distributions. The principal redraws $m$ conditional on $(\th', \tau)$. The agent redraws $\th'$ conditional on $(\th, m)$. To construct these conditional distributions, apply disintegration of measures \citep[Theorem 1.25, p.~39]{Kallenberg2017}. This result applies to Borel probability measures, so we first restrict our measures to the Borel $\s$-algebra, then apply the theorem, and finally extend the resulting measures to the universal completion.}

\subsection{Proof of \texorpdfstring{\cref{res:replacement}}{Theorem \ref{res:replacement}}}

Let $f$ be a social choice function that is canonically implemented by a mechanism $(t,g)$ in which $t(\th) = \psi$. Define the mechanism $(t',g')$
to coincide with $(t,g)$ except for the following modifications. Set $t'(\th) = \tau$. Choose $k_1$ and $k_0$ from the definition of $\tau \succeq_\th \psi$. For each $s = 0,1$, set
\[
    g' (\th,\tau,s) = k_s g (\th,\psi,1) + (1 - k_s) g (\th,\psi,0) \in \D(X).
\]
Under the mechanism $(t',g')$, if type $\th'$ reports type $\th$ and tries on test $\tau$ with probability $a$, the resulting decision will be
 \[
    p(a|\th') g(\th,\psi,1) + (1 - p(a|\th'))g(\th, \psi, 0) \in \D(X),
\]
where
\[
    p(a|\th') =  a \Brac{ \pi (\tau | \th') k_1  + (1 - \pi (\tau | \th')) k_0} + (1 - a) k_0. 
\]
From the definition of $\tau \succeq_\th \psi$, we have $p(a|\th') \leq \pi(\psi |\th')$ for all $a$ in $[0,1]$ and all types $\th'$, with equality if $a = 1$ and $\th' = \th$. Therefore, $(t', g')$ replicates the social choice function $f$ without introducing any new deviation outcomes for any type.

\subsection{Proof of \texorpdfstring{\cref{res:test_uniqueness}}{Proposition \ref{res:test_uniqueness}}} \label{sec:proof_test_uniqueness}

Fix a type $\th$ and tests $\tau_1$ and $\tau_2$.

One direction is clear. If $\pi(\tau_1 | \cdot) = \pi (\tau_2 | \cdot)$, then we can set  $(k_0, k_1) = (0,1)$ to see that $\tau_1 \succeq_\th \tau_2$ and $\tau_2 \succeq_\th \tau_1$. If $\th$ is minimal on test $\tau_1$ then we see that $\tau_2 \succeq_\th \tau_1$ by setting $k_0 = k_1 = \pi ( \tau_1 | \th)$. Symmetrically, if $\th$ is minimal on $\tau_2$, then we see that $\tau_1 \succeq_\th \tau_2$ by setting $k_0 = k_1 = \pi (\tau_2 | \th)$.

For the converse, assume $\tau_1 \sim_{\th} \tau_2$. Choose $(k_0, k_1)$ from the definition of $\tau_1 \succeq_\th \tau_2$ and $(k_0', k_1')$ from the definition of $\tau_2 \succeq_\th \tau_1$. Suppose type $\th$ is not minimal on one of the tests, say $\tau_1$. Hence there exists $\th'$ such that $\pi(\tau_1 |\th) > \pi (\tau_1| \th')$.  We prove that $\pi (\tau_1 | \cdot) = \pi( \tau_2 | \cdot)$. 

We use Markov transition notation; see \cref{sec:nonbinary}. Let $k$ and $k'$ denote the Markov transitions associated with $(k_0, k_1)$ and $(k_0', k_1')$, respectively. Let $\pi_{\tau | \th}$ denote the probability measure that puts probability $\pi(\tau | \th)$ on $s = 1$. We have
\[
    \pi_{\tau_1| \th} k k' = \pi_{\tau_2|\th} k' = \pi_{\tau_1 | \th}
    \qquad
    \text{and}
    \qquad
    \pi_{\tau_1|\th'} k k' \preceq_{\ST} \pi_{\tau_2 | \th'} k' \preceq_{\ST} \pi_{\tau_1 | \th'}.
\]
In terms of the probability on $s = 1$, we can express this system as
\begin{equation*}
\begin{aligned}
       k_0 k_1' + (1 - k_0)k_0' + \pi ( \tau_1 | \th) (k_1 - k_0)(k_1' - k_0') &= \pi(\tau_1 | \th), \\
   k_0 k_1' + (1 - k_0)k_0' + \pi ( \tau_1 | \th') (k_1 - k_0)(k_1' - k_0') &\leq \pi(\tau_1 | \th').
\end{aligned}
\end{equation*}
 After subtracting, we conclude that 
 \[
    [\pi(\tau_1 | \th)- \pi (\tau_1 | \th')] (k_1 - k_0)(k_1' - k_0') \geq \pi(\tau_1 | \th) - \pi (\tau_1 |\th').
\]
Since $\pi(\tau_1 | \th)- \pi (\tau_1 | \th') > 0$, it follows that $(k_0, k_1) = (k_0', k_1') = (0,1)$, and hence $\pi(\tau_1 | \cdot) = \pi(\tau_2 | \cdot)$. 

\subsection{Proof of \texorpdfstring{\cref{res:implementation}}{Theorem \ref{res:implementation}}}

$\eqref{it:stat} \implies \eqref{it:decision}$. Let $\hat{t}$ be a most-discerning testing function. For each type $\th$ and test $\psi$, select probabilities $k_0 (\th, \psi)$ and $k_1 (\th, \psi)$ satisfying the definition of $\hat{t}(\th) \succeq_\th \psi$; \cref{sec:measurable_selection_score_conversion}
shows that there exists a \emph{measurable} selection. Fix a decision environment $(X, u)$. Let $f$ be an implementable social choice function. By the revelation principle (\cref{res:revelation_principle}), $f$ is canonically implemented by some mechanism $(t, g)$. Consider a mechanism $(\hat{t}, \hat{g})$ in which $\hat{g}$ satisfies
\begin{equation*} 
    \hat{g}( \th, \hat{t}(\th), s) = \E_{\psi \sim t(\th)} \Brac{ k_s ( \th, \psi) g( \th, \psi, 1) + (1 - k_s ( \th, \psi)) g( \th, \psi, 0)} \in \D(X),
\end{equation*}
for all types $\th$ and scores $s = 0,1$.\footnote{In a slight abuse of notation, $t$ maps $\Th$ into $\D(T)$, while $\hat{t}$ maps $\Th$ into $T$.}

Under the mechanism $(\hat{t}, \hat{g})$, if type $\th'$ reports type $\th$ and then tries on test $\hat{t}(\th)$ with probability $a$, the resulting decision will be
 \[
    \E_{\psi \sim t(\th)} \Brac{ p(a, \th, \psi |\th') g( \th, \psi, 1) + (1- p(a,\th, \psi|\th')) g( \th, \psi, 0) } \in \D(X),
 \]
 where 
 \[
    p(a, \th, \psi| \th') =  a \Brac{ \pi (\hat{t}(\th) | \th') k_1(\th, \psi)  + (1 - \pi (\hat{t}(\th) | \th')) k_0 (\th, \psi)} + (1 - a) k_0 (\th, \psi). 
 \]
For each type $\th$ and test $\psi$, the definition of $\hat{t}(\th) \succeq_\th \psi$ guarantees that $p(a, \th, \psi |\th') \leq \pi(\psi |\th')$ for all $a$ in $[0,1]$ and all types $\th'$, with equality if $a = 1$ and $\th' = \th$. Therefore, $(\hat{t}, \hat{g})$ replicates the social choice function $f$ without introducing any new deviation outcomes for any type.\footnote{Our argument is similar in spirit to  \citeapos{DeOliveira2018} elegant proof of Blackwell's theorem using diagrams.}  

$\eqref{it:decision} \implies \eqref{it:stat}$.  Fix a type $\th$ and a test $\psi$. We will prove that $\hat{t}(\th) \succeq_\th \psi$. 

Construct a decision environment $(X,u)$ as follows. The decision set $X$ consists of three decisions, denoted $\bar{x}$, $\ubar{x}$, and $y$. Every type gets utility $1$ from decision $\bar{x}$ and utility $0$ from decision $\ubar{x}$. Each type $\th'$ gets utility $\pi(\psi | \th')$ from decision $y$. 

Consider the following mechanism. If the agent reports $\th'$ with $\th' \neq \th$, the principal selects $y$ (the test and score do not matter). If the agent reports $\th$, the principal gives test $\psi$ and then selects $\bar{x}$ if the agent passes and $\ubar{x}$ if the agent fails. Observe that truth-telling and trying is a best response for every type. Denote the induced social choice function by $f$. 

By \eqref{it:decision}, $f$ can be canonically implemented by $(\hat{t}, \hat{g})$, for some outcome rule $\hat{g}$.  For $s = 0,1$, let $k_s$ be the probability that $\hat{g}(\th, \hat{t}(\th), s)$ assigns to $\bar{x}$. We must have $k_1\geq k_0$; otherwise, type $\th$ could profitably deviate by intentionally failing test $\hat{t}(\th)$.\footnote{If $\pi (\hat{t}(\th)| \th) > 0$, this holds because  $\hat{g}(\th, \hat{t} (\th) , 1)$ must concentrate on $\{ \ubar{x}, \bar{x}\}$, and type $\th$ weakly prefers $y$ to $\ubar{x}$. If $\pi( \hat{t}(\th) | \th) = 0$, then we may assume $k_1 = k_0$ since implementation is preserved by redefining $\hat{g}(\th, \hat{t}(\th), 1)$ to equal $\hat{g}(\th, \hat{t}(\th), 0)$.} Since this mechanism implements $f$, the probabilities $k_0$ and $k_1$ satisfy \eqref{it:theta_equality} in the definition of $\hat{t}(\th) \succeq_\th \psi$. Since no type $\th'$ can profit from reporting $\th$ and trying on test $\hat{t}(\th)$, we get \eqref{it:theta_prime_inequality}. Therefore, $\hat{t}(\th) \succeq_\th \psi$, as desired. 

\subsection{Proof of \texorpdfstring{\cref{res:relative_performance_tests}}{Proposition \ref{res:relative_performance_tests}}}

There are two cases. 

\begin{enumerate}
    \item Suppose $\pi (\tau | \th) \geq \pi ( \psi | \th) > 0$. If \eqref{eq:char_tau_easy} holds, then \cref{def:discernment_order} is satisfied with
\[
    k_0 = 0 
    \quad
    \text{and} 
    \quad 
    k_1 = \frac{\pi (\psi | \th)}{\pi(\tau | \th)}.
\]

\item  Suppose $\pi (\tau | \th) \leq \pi (\psi | \th) < 1$. If \eqref{eq:char_tau_hard} holds, then \cref{def:discernment_order} is satisfied with
\[
    k_0 = \frac{\pi (\psi | \th) - \pi ( \tau | \th)}{1 - \pi (\tau | \th)} \quad \text{and} \quad 
     k_1 = 1. 
\]
To see this, multiply each side of \eqref{eq:char_tau_hard}  by 
$1 - \pi (\psi | \th)$. 
Subtract each side of the resulting inequality from $1$ (and flip the direction of the inequality). 
\end{enumerate}

\subsection{Proof of \texorpdfstring{\cref{res:most_discerning_authentication}}{Proposition \ref{res:most_discerning_authentication}}}
Let $\a$ be an authentication rate satisfying $\a ( \th | \th) \geq \max \{ \a ( \th' | \th), \a(\th | \th') \}$ for all $\th, \th' \in \Th$. First, observe that \eqref{eq:alpha_NRC} is trivially satisfied if $\th_1 = \th_2$ or $\th_2 = \th_3$. Fix distinct types $\th_2$ and $\th_3$. By \cref{rem:authentication_linear_system}, it suffices to show that $\tau_{\th_2}^\alpha \succeq_{\th_2} \tau_{\th_3}^{\alpha}$ if and only if 
\begin{equation} \label{eq:alpha_cond}
    \a  (\th_3 | \th_2) \a(\th_2 | \th_1) \leq \a (\th_3 | \th_1) \a (\th_2 |\th_2), \quad \text{for all}~\th_1 \in \Th \setminus \{ \th_2\}. 
\end{equation}
There are two cases.

\begin{enumerate}
    \item Suppose $\a (\th_2 | \th_2) = 0$. It follows from the assumption on $\a$ that $\a (\th_3 | \th_2) = 0$. Thus, \eqref{eq:alpha_cond} is satisfied (because both sides are zero). Also, $\tau_{\th_2}^\alpha \succeq_{\th_2} \tau_{\th_3}^\alpha$ because the system \eqref{eq:alpha_system_main_text}, with $\th = \th_2$ and $\th' = \th_3$, is solved by $k_0 = 0$ and $k_1 = 1$ (by the assumption on $\a$).
    \item Suppose $\a (\th_2 | \th_2) > 0$. If \eqref{eq:alpha_cond} holds, then $\tau_{\th_2}^\alpha \succeq_{\th_2} \tau_{\th_3}^{\alpha}$ because the system \eqref{eq:alpha_system_main_text}, with $\th = \th_2$ and $\th' = \th_3$, is solved by $k_0 = 0$ and $k_1 = \a (\th_3 | \th_2)/\a (\th_2 | \th_2)$; note that $\a (\th_3 | \th_2)/\a (\th_2 | \th_2) \leq 1$ by the assumption on $\a$. Conversely, if $\tau_{\th_2}^\a \succeq_{\th_2} \tau_{\th_3}^\a$, then the system \eqref{eq:alpha_system_main_text}, with $\th = \th_2$ and $\th' = \th_3$, has a nonnegative solution $(k_0, k_1)$. We claim that this system is also solved by 
    \[
        k_0' = 0 \quad \text{and} \quad k_1' = k_1 + \frac{ 1 - \a (\th_2 | \th_2)}{\a ( \th_2| \th_2)} k_0.
    \]
    To see this, note that this modification leaves the equality in \eqref{eq:alpha_system_main_text} unchanged and changes the left side of the $\th''$-inequality by 
    \[
        \Brac{(1 - \a (\th_2 | \th_2)) \frac{ \a ( \th_2 | \th'')}{\a ( \th_2 | \th_2) } - (1  - \a (\th_2 | \th''))} k_0,
    \]
    which is nonpositive because $\a ( \th_2 | \th_2) \geq \a( \th_2 | \th'')$, by the assumption on $\a$. Now examine the new solution $(k_0', k_1')$ of the system \eqref{eq:alpha_system_main_text}, with $\th = \th_2$ and $\th' = \th_3$. Since $k_0' = 0$, the equality gives $k_1' = \a ( \th_3 | \th_2) / \a ( \th_2 | \th_2)$. In each inequality, scale each side by $\a ( \th_2 | \th_2)$ to get \eqref{eq:alpha_cond}, as desired. 
    
\end{enumerate}

\subsection{Proof of \texorpdfstring{\cref{res:nonlinear_optimum}}{Proposition \ref{res:nonlinear_optimum}}} \label{sec:proof_nonlinear}

The following preliminary lemma is proven in \cref{sec:proof_bounded_quantity}. 

\begin{lem}[Bounded mechanisms] \label{res:bounded_quantity}
Let $(q,t)$ be an incentive compatible mechanism. There exists a bounded, incentive compatible mechanism $(\bar{q}, \bar{t})$ such that either
\begin{enumerate*}[label = (\roman*), ref = \roman*] \item \label{it:ae} $(\bar{q}, \bar{t})$ and $(q,t)$ agree almost surely, or \item \label{it:improve} the principal strictly prefers $(\bar{q}, \bar{t})$ to $(q,t)$.
\end{enumerate*}
\end{lem}

By \cref{res:bounded_quantity}, it suffices to prove that $(q^\ast, t^\ast)$ is the essentially unique optimum among all bounded, incentive compatible mechanisms $(q,t)$. By setting $U(\th) = \th q(\th) - t(\th)$, we can equivalently specify a mechanism $(q,t)$ as a quantity--utility pair $(q,U)$. Note that $(q,t)$ is bounded if and only if $(q,U)$ is. 

\begin{lem}[Envelope theorem bound] \label{lem:utility_bound}
Let $(q,U)$ be a quantity--utility pair. If $(q,U)$ is bounded and incentive compatible, then for each type $\th$, we have
\begin{equation} \label{eq:U_pointwise_bound}
	U(\th) \geq \int_{\ubar{\th}}^{\th} \a ( \xi | \th ) q(\xi) \de \xi.
\end{equation}
\end{lem}

\cref{lem:utility_bound} is proven in \cref{sec:proof_envelope}. We turn to the main proof of \cref{res:nonlinear_optimum}. Let $(q, U)$ be a bounded, incentive compatible quantity--utility pair. We can bound the principal's objective by applying \cref{lem:utility_bound} and then switching the order of integration:
\begin{equation*}
	\int_{\ubar{\th}}^{\bar{\th}} [ \th q(\th) - c(q(\th)) - U(\th)] f(\th) \de \th  \leq \int_{\ubar{\th}}^{\bar{\th}} [ \varphi(\th) q(\th) - c(q(\th))] f(\th) \de \th,
\end{equation*}
with equality if and only if \eqref{eq:U_pointwise_bound} holds with equality for almost every type $\th$. For each type $\th$, the integrand in brackets on the right side is uniquely maximized by $q^\ast(\th)$. The transfer function $t^\ast$ ensures that $U$ satisfies \eqref{eq:U_pointwise_bound} with equality for every type $\th$. 

To complete the proof, we check that $(q^\ast, t^\ast)$ satisfies global incentive compatibility if the quantity function $q^\ast$ satisfies the following monotonicity condition: Whenever $\ubar{\th} \leq \xi_1 \leq \xi_2 \leq \th$, we have 
\begin{equation} \label{eq:monotone}
    \a (\xi_1 | \th) q^\ast (\xi_1) \leq \a(\xi_2 | \th) q^\ast (\xi_2).
\end{equation}
This monotonicity condition holds because $q^\ast$ is weakly increasing (since $\varphi$ is weakly increasing).

The global incentive constraints require that for all types $\th$ and $\th'$, we have
\begin{equation*}
    U(\th) \geq \a (\th' | \th) [ U(\th') + (\th - \th') q^\ast(\th')],
\end{equation*}
or equivalently,
\begin{equation} \label{eq:weak_ineq}
    U(\th) - \a (\th' | \th) U(\th') \geq (\th - \th') \a ( \th' | \th) q^\ast (\th').
\end{equation}
Plug in the right side of \eqref{eq:U_pointwise_bound} for $U$ to get the condition
\begin{equation} \label{eq:global_IC_fixed}
     \int_{\ubar{\th}}^{\th} \a (\xi | \th) q^\ast (\xi ) \de \xi  -     \int_{\ubar{\th}}^{\th'} \a (\xi | \th') \a(\th' | \th) q^\ast (\xi) \de \xi \geq (\th - \th') \a (\th' | \th) q^\ast(\th').
\end{equation}
We separate into cases. If $\th > \th'$, then \eqref{eq:global_IC_fixed} is equivalent to  
\[
    \int_{\th'}^{\th} \a( \xi | \th) q^\ast (\xi) \geq ( \th - \th') \a (\th' | \th) q^\ast(\th').
\]
If $\th < \th'$, then \eqref{eq:global_IC_fixed} holds if 
\[
    \int_{\th}^{\th'} \a( \xi | \th') q^\ast (\xi) \leq ( \th' - \th) q^\ast(\th').
\]
In each case, the inequality is guaranteed by the monotonicity condition in \eqref{eq:monotone}.

\subsection{Proof of \texorpdfstring{\cref{res:posted_price}}{Proposition \ref{res:posted_price}}} \label{sec:proof_posted_price}

We follow the proof of \cref{res:nonlinear_optimum} in \cref{sec:proof_nonlinear}. As before, it suffices to prove essentially unique optimality among all bounded, incentive compatible mechanisms.\footnote{Any quantity function $q \colon \Th \to [0,1]$ is bounded. By \cref{res:bounded_quantity}, it suffices to consider bounded transfer functions.} For any bounded, incentive compatible quantity--utility pair $(q,U)$, we have
\[
	\int_{\ubar{\th}}^{\bar{\th}} [ \th q(\th) - c q(\th) - U(\th)] f(\th) \de \th  \leq \int_{\ubar{\th}}^{\bar{\th}} (\varphi(\th) -c ) q(\th)  f(\th) \de \th,
\]
with equality if and only if \eqref{eq:U_pointwise_bound} holds with equality for almost every type $\th$. For each type $\th$, the integrand on the right side is maximized by $q^\ast(\th)$, uniquely so if $\th \neq \th^\ast$. The transfer function $t^\ast$ ensures that $U$ satisfies \eqref{eq:U_pointwise_bound} with equality for every type $\th$. 

To check that $(q^\ast, t^\ast)$ is globally incentive compatible, follow the argument from the proof of \cref{res:nonlinear_optimum} in \cref{sec:proof_nonlinear}.

\subsection{Proof of Lemma~\ref{res:bounded_quantity}} \label{sec:proof_bounded_quantity}

Let $(q,t)$ be an incentive compatible mechanism. Let
\[
    \Th_0 = \{ \th \in \Th : t(\th) - c(q(\th)) \geq 0 \}.
\]
Since $\lim_{q \to \infty} c'(q) > \bar{\th}$, we may choose $L$ such that $\bar{\th} q - c(q) < 0$ for all $q > L$. 
For all $\th \in \Th_0$, it follows from the participation constraint that $\th q(\th) - c(q(\th)) \geq 0$, so $q(\th) \leq L$ and hence $0 \leq t(\th) \leq \bar{\th} L$. For each type $\th$, let $\varphi(\th)$ be the closure of the bounded set
\[
    \Set{ \Paren{ \a ( \th' | \th) q(\th'), \a ( \th' | \th) t(\th')} : \th' \in \Th_0 }.
\]
By the measurable maximum theorem \cite[18.19, p.~605]{AliprantisBorder2006}, the correspondence \[
    \th \mapsto \argmax_{(q',t') \in \varphi (\th)} ( \th q ' - t')
\]
admits a measurable selection $(\tilde{q},\tilde{t}) \colon \Th \to [0,L] \times [0, \bar{\th} L]$. Define $(\bar{q},\bar{t})$ to equal $(q,t)$ on $\Th_0$ and $(\tilde{q}, \tilde{t})$ on $\Th \setminus \Th_0$.

By the supermultiplicativity of $\a$ (see \cref{res:most_discerning_authentication}), we have
\[
    \a ( \th'' | \th) \geq \a ( \th''| \th') \a ( \th' | \th),
\]
for all types $\th \in \Th$ and all reports $\th' \in \Th \setminus \Th_0$ and $\th'' \in \Th_0$. Thus, it can be checked that $(\bar{q}, \bar{t})$ is incentive compatible. 

Now we complete the proof. If $\Th \setminus \Th_0$ has measure zero, we get \eqref{it:ae}. If $\Th \setminus \Th_0$ has positive measure, we claim that
\eqref{it:improve} holds. We show that for each $\th \in \Th \setminus \Th_0$, the principal strictly prefers $(\tilde{q}(\th), \tilde{t}(\th))$ to $(q(\th), t(\th))$. Fix $\th \in \Th \setminus \Th_0$. For each
 $\th' \in \Th_0$, we have
\[
    \a (\th' | \th) t(\th') - c \Paren{ \a (\th' | \th) q(\th')}
    \geq 
    \a (\th' | \th) t(\th') - \a (\th' | \th) c (q(\th')) 
    \geq 0,
\]
where the first inequality uses the convexity of $c$. We conclude that
\[
    \tilde{t}(\th) - c(\tilde{q}(\th)) \geq 0 > t(\th) - c(q(\th)).
\]

\subsection{Proof of \texorpdfstring{\cref{lem:utility_bound}}{Lemma \ref{lem:utility_bound}}} \label{sec:proof_envelope}

Let $(q,U)$ be a bounded, incentive compatible quantity--utility pair. We first check that $U$ is absolutely continuous.  Choose $\th$ and $\th'$ such that $U(\th') \geq U(\th)$. By incentive compatibility,
\[
    U(\th) \geq \a ( \th' | \th) \Brac{ U(\th') + (\th - \th') q ( \th')}.
\]
Therefore,
\begin{equation*}
\begin{aligned}
    0 &\leq U(\th') - U(\th) \\
     &\leq  (1 - \a (\th' | \th))U(\th') + \a(\th' | \th) (\th' - \th)   q(\th') \\
     &\leq (1- \a ( \th' | \th))  \| U \|_{\infty} + |\th' - \th| \cdot \|q\|_{\infty}.
\end{aligned}
\end{equation*}
Since $1 - e^{-x} \leq x$, it follows that
\[
   0 \leq  U(\th') - U(\th) \leq C \Abs{ \int_{\th'}^{\th} (\l(\xi) + 1) \de \xi},
\]
where $C = \max\{ \| U\|_{\infty}, \|q\|_{\infty} \}$. Since $\l + 1$ is integrable over $[\ubar{\th}, \bar{\th}]$, we conclude that $U$ is absolutely continuous.

Now we prove \eqref{eq:U_pointwise_bound}. Define the auxiliary function $\D$ on $[\ubar{\th}, \bar{\th}]$ by
\begin{equation*}
\begin{aligned}
    \D ( \th) 
    &= 
    \a(\th | \bar{\th}) \biggl( U (\th) - \int_{\ubar{\th}}^{\th} \a (\xi | \th) q(\xi) \de \xi \biggr) \\
    &= \a ( \th | \bar{\th}) U(\th) - \int_{\ubar{\th}}^{\th} \a(\xi | \bar{\th}) q(\xi) \de \xi.
\end{aligned}
\end{equation*}
We prove that $\D$ is nonnegative. The function $\D$ is absolutely continuous since it is the product of absolutely continuous functions. Let $u(\th' | \th) = \a(\th' | \th) [ \th q(\th') - t(\th')]$. By Theorem 1 in \cite{MilgromSegal2002}, whenever $U$ is differentiable, we have
\[
	U'(\th) \geq D_{2+} u(\th | \th) = 	q(\th) - \l(\th) U(\th),
\]
where $D_{2+} u(\th|\th)$ denotes the right derivative with respect to the second argument.\footnote{That is, $D_{2+} u(\th|\th) = \lim_{h \downarrow 0} h^{-1} (u(\th|\th + h) - u(\th|\th))$.}
Let $I(\th) = \int_{\ubar{\th}}^{\th} \a ( \xi | \bar{\th}) q(\xi) \de \xi$. At almost every $\th$ in $[\ubar{\th}, \bar{\th}]$, the absolutely continuous functions functions $\D$, $U$, $\a (\cdot | \bar{\th})$, and $I$ are all differentiable, so we get 
\begin{equation*}
\begin{aligned}
	\D'(\th) 
	&= \l(\th) \a (\th| \bar{\th}) U(\th) + \a (\th | \bar{\th}) U'(\th) - \a ( \th | \bar{\th}) q(\th) \\
	&= \a (\th | \bar{\th}) \Brac{  U'(\th) - ( q(\th) - \l (\th) U(\th) ) } \\
	& \geq 0. 
\end{aligned}
\end{equation*}
 By the fundamental theorem of calculus, for $\ubar{\th} \leq \th \leq \bar{\th}$, we have
 \[
 \D (\th) \geq \D (\ubar{\th}) = U(\ubar{\th}) \geq 0,
\]
where the last inequality follows from the participation constraint. 

\newpage
\bibliographystyle{aer}
\bibliography{newest_lit}

\newpage
\section{Online appendix} \label{sec:additional_results}

\subsection{Insufficiency of mixed strategies in reduced form} \label{sec:mixing_randomization}

Consider a modification of \cref{sec:cautionary-example}. The type space and the verification technology are as in \cref{fig:challenge}. Now there are three allocations---nothing, low-quality, and high-quality---with associated type-independent utilities $0$, $u_\ell$, and $u_h$. Assume $0 < u_\ell < u_h$ and $u_\ell \geq u_h /2$. Consider the social choice function that allocates the high-quality good to type $\th_1$ and the low-quality good to types $\th_2$ and $\th_3$. 

We claim that this social choice function cannot be implemented in the reduced-form model, even if the agent uses a mixed strategy. Type $\th_1$ can pass only as type $\th_1$ or as type $\th_2$. So for some $\th \in \{ \th_1, \th_2\}$, the principal must give the high-quality good to the agent if he passes as type $\th$. But type $\th_3$ can pass as type $\th_1$, and type $\th_2$ can pass as type $\th_2$, so at least one of the types $\th_3$ and $\th_2$ has a strictly profitable deviation.

Now replace the authentication rate with a testing technology consisting of three tests, denoted $\tau_1$, $\tau_2$, $\tau_3$. Test $\tau_i$ can be passed by those types that can pass as $\th_i$ in the reduced-form model. In this testing model, the principal can implement the specified social choice function. If the agent reports type $\th_2$ or $\th_3$, he is given the low-quality good. If the agent reports type $\th_1$, then the principal gives either test $\tau_1$ or test $\tau_2$, each with probability $1/2$. Whichever test is given, the agent gets the high-quality good if he passes and nothing if he fails. If either type $\th_2$ or type $\th_3$ deviates by reporting $\th_1$, then he gets the high-quality good with probability at most $1/2$, and otherwise he gets nothing. This deviation is unprofitable since $u_\ell \geq u_h /2$.

\subsection{Most-discerning correspondences} \label{sec:most-discerning_correspondences}

 Even if the testing technology does not admit a most-discerning testing function, we can still use the discernment orders to reduce the class of tests that need to be considered.

\begin{defn}[Most-discerning correspondence]  A subset $T_0$ of $T$ is \emph{most $\th$-discerning} if for each test $\psi$ in $T$ there exists a test $\tau$ in  $T_0$ such that $\tau \succeq_\th \psi$. A correspondence $\hat{T} \colon \Th \twoheadrightarrow T$ is \emph{most-discerning} if for each type $\th$ the set $\hat{T}(\th)$ is most $\th$-discerning.
\end{defn}

A testing rule $\hat{t} \colon \Th \to \D (T)$ is \emph{supported on} a correspondence $\hat{T} \colon \Th \twoheadrightarrow T$ if $\supp \hat{t} (\th) \subset \hat{T}(\th)$ for each type $\th$. The next result says that if a correspondence is most-discerning, then we can restrict attention to testing rules supported on that correspondence. 

\begin{thm}[Implementation with a most-discerning correspondence] \label{res:implementation_sharp}
Suppose that the passage rate $\pi$ is continuous. Let\/ $\hat{T}$ be a weakly measurable\footnote{That is, the lower inverse $\{ \th  \in \Th : T(\th) \cap G  \neq \varnothing \}$ is universally measurable for each open subset $G$ of $T$; see \citet[p~592]{AliprantisBorder2006}.} correspondence from $\Th$ to $T$ with closed values. If\/ $\hat{T}$ is most-discerning, then for every implementable social choice function $f$, there exists a testing rule $\hat{t}$ supported on $\hat{T}$ such that $f$ is canonically implementable with $\hat{t}$. 
\end{thm}

The proof is essentially the same as the proof of \cref{res:implementation}. For each type $\th$ and test $\psi$, there exists a test $\tau$ in $\hat{T}(\th)$ such that $\tau \succeq_\th \psi$. But we must check that there exists such a selection that is measurable; see \cref{sec:measurable_selection_score_conversion}. The regularity conditions on $\pi$ and $\hat{T}$ ensure that a measurable selection exists. If we can independently construct a measurable selection, then these conditions are not needed. 

\subsection{Beyond exponential authentication rates} \label{sec:beyond_exponential}

Suppose that the verification technology is represented by a Borel measurable, most-discerning authentication rate $\a \colon \Th \times \Th \to [0,1]$ that satisfies the following conditions.
\begin{enumerate} [label = (\roman*), ref = \roman*]
	\item \label{it:no_false_detection} $\a (\th |  \th) = 1$ for all types $\th$.
	\item \label{it:AC} For each type $\th'$, the function $\th \mapsto \a (\th' | \th)$ is absolutely continuous.
	\item \label{it:integrable} For each type $\th$, the right and left partial derivatives (with respect to the second argument) $D_{2+} \a (\th|\th)$ and $D_{2-} \a (\th| \th)$ exist, and the functions $\th \mapsto D_{2+} \a (\th|\th)$ and $\th \mapsto D_{2-} \a (\th|\th)$ are integrable.
\end{enumerate}
Condition \ref{it:no_false_detection} ensures that the agent is authenticated if he reports truthfully. Conditions \ref{it:AC} and \ref{it:integrable} allow us to apply the envelope theorem. In particular, the exponential authentication rates studied in the main text satisfy these assumptions. 

Define the right and left local precision functions $\l_+, \l_- \colon \Th \to \R_+$ by
\begin{equation} \label{eq:precision_fuction}
    \l_+ (\th) = -D_{2+} \a ( \th | \th), 
    \qquad
    \l_- (\th) = D_{2-} \a (\th, \th). 
\end{equation}
Define the function $\L$ by 
\[
	\L (\th'| \th)
	=
	\begin{cases} 
		 \exp \Paren{ - \int_{\th'}^{\th} \l_+ (\xi) \de \xi } &\text{if } \th \geq \th', \\[10pt]
		\exp \Paren{ - \int_{\th}^{\th'} \l_- (\xi) \de \xi } &\text{if } \th < \th'.
	\end{cases}
\]
The function $\L$ is determined only by the local behavior of $\a$ near the diagonal.

\begin{lem}[Lower bound on authentication rate] \label{res:lower_bound} 
For all types $\th$ and $\th'$, we have $\a (\th' | \th) \geq \L (\th' | \th)$.
\end{lem}

\cref{res:lower_bound} is proven in \cref{sec:proof_lower_bound}. For the exponential authentication rate $\a$ considered in the main text, we have $\l_+(\th) = \l_-(\th) = \l(\th)$, so $\a ( \th' | \th) = \L (\th' | \th)$ for all types $\th$ and $\th'$. Therefore, among all most-discerning authentication rates satisfying \eqref{it:no_false_detection}--\eqref{it:integrable} with $- D_{2+} \a(\th|\th) = D_{2-} \a(\th | \th) = \l(\th)$ for each $\th$, the exponential authentication rate with precision function $\l$ makes the global incentive constraints weakest. 

In this general setting,  we show under further regularity conditions that the optimal mechanisms take the same form, except that the virtual value $\varphi$ is defined with $\L$ in place of $\a$:
\[
    \varphi(\th) = \th - \frac{1}{f(\th)} \int_{\th}^{\bar{\th}} \L ( \th| \xi) f(\xi) \de \xi.
\]
\cref{res:bounded_quantity} goes through with exactly the same proof. \cref{lem:utility_bound} can be shown to  hold with $\L (\xi | \th)$ in place of $\a ( \xi | \th)$.%
\footnote{The proof is similar to the proof in \cref{sec:proof_envelope}. To establish absolute continuity, apply \cref{res:lower_bound} and put $\l_+ \vee \l_-$ in place of $\l$. To establish the bound, use $\L$ in place of $\a$ in the definition of the auxiliary function $\D$. The rest of the proof goes through with $\l_+$ in place of $\l$.} Therefore, \cref{res:nonlinear_optimum} and 
\cref{res:posted_price} go through, with the redefined virtual value,  if 
\begin{enumerate*}[label = (\alph*), ref = \alph*]
    \item \label{it:mon} the monotonicity condition \eqref{eq:monotone} holds with $\L$ in place of $\a$, and \item \label{it:global}  the following global bound is satisfied:
\end{enumerate*} For $\th > \th'$, 
\begin{equation} \label{eq:global_bound}
    \a ( \th'|\th) \leq \L (\th' | \th) \frac{ \int_{\ubar{\th}}^{\th'} \L (\xi | \th) q^\ast(\xi) \de \xi + \int_{\th'}^{\th} \L (\xi | \th) q^\ast(\xi) \de \xi}{\int_{\ubar{\th}}^{\th'} \L (\xi | \th) q^\ast(\xi) \de \xi + \int_{\th'}^{\th} \L ( \th' | \th) q^\ast(\th') \de \xi}.
\end{equation}
Intuitively, the more rapidly the map $\xi \mapsto \L(\xi | \th) q^\ast(\xi)$ increases over the interval $[0, \th]$, the more slack there is for $\a ( \th' | \th)$ to increase above $\L(\th'| \th)$.

We check that \eqref{it:mon} and \eqref{it:global} imply global incentive compatibility. By \cref{lem:utility_bound}, the analogue of \eqref{eq:global_IC_fixed} is 
\begin{equation} \label{eq:IC_gen}
    \int_{\ubar{\th}}^{\th} \L (\xi | \th) q^\ast (\xi) \de \xi - \int_{\ubar{\th}}^{\th'} \L (\xi | \th') \a (\th' | \th) q^\ast (\xi) \de \xi \geq (\th - \th') \a (\th' | \th) q^\ast (\th').
\end{equation}
We separate into cases. If $\th < \th'$, then \eqref{eq:IC_gen} holds if 
\[
    \int_{\th}^{\th'} \L (\xi | \th') q^\ast ( \xi) \de \xi \leq ( \th' - \th) q^\ast (\th'),
\]
which is guaranteed by \eqref{it:mon}. If $\th > \th'$, then \eqref{eq:IC_gen} is equivalent to 
\[
    \int_{\ubar{\th}}^{\th} \L (\xi | \th) q^\ast (\xi)  \de \xi
    \geq \frac{\a(\th' | \th)}{\L(\th' | \th)} \Brac{ \int_{\ubar{\th}}^{\th'} \L(\xi | \th) q^\ast(\xi) \de \xi + (\th - \th') \L (\th' | \th) q^\ast(\th')}.
\]
Rearranging, we see that this inequality is equivalent to \eqref{eq:global_bound}.

\subsection{Nonbinary tests} \label{sec:nonbinary_appendix}

First, we check that $\succeq_\th$ is reflexive and transitive. Reflexivity is immediate by taking $k$ to be the identity, which maps each score $s$ to the point mass $\d_s$. For transitivity, it follows from \citet[Proposition 1, pp.~901--902]{KamaeEtal1977} that (i) the order $\succeq_{\ST}$ is preserved by increasing Markov transitions; and (ii) the composition $k_1 k_2 \colon S \to \D(S)$ defined by $(k_1 k_2)(s'| s) = \sum_{s''} k_2(s' | s'') k_1 ( s'' |s)$ is increasing if $k_1$ and $k_2$ are increasing.

In the main model, if type $\th$ tries on test $\tau$ with probability $a$, he passes with probability $a \pi (\tau | \th)$. Therefore, type $\th$ can achieve on test $\tau$ any passage probability $p$ satisfying $p \leq \pi(\tau | \th)$.  In the general case, on a nonbinary test $\tau$, type $\th$ chooses a Markov transition $d \colon S \to \D(S)$ that is \emph{downward} in the sense that $d(s' | s) = 0$ unless $s \succeq s'$. Then Nature draws the score from the distribution $\pi_{\tau | \th} d$. By \citet[Theorem 1, p.~900]{KamaeEtal1977}, type $\th$ can achieve on test $\tau$ a score distribution $p$ in $\D(S)$ if and only if $p \preceq_{\ST} \pi_{\tau | \th}$. Given a general mechanism $(M, M'; t, r', g)$, a strategy for the agent is a pair $(r,d)$ consisting of a messaging strategy $r \colon \Th \to \D(M)$ and an action strategy $d \colon \Th \times M \times T \times M' \times S \to \D(S)$ such that $d_{\th, m, \tau, m'} \colon S \to \D(S)$ is downward for each $(\th, m, \tau, m') \in \Th \times M \times T \times M'$.
 
In this setting with nonbinary tests, the following results go through: the revelation principle (\cref{res:revelation_principle}), the replacement theorem (\cref{res:replacement}), and the forward implication in the main implementation theorem (\cref{res:implementation}).
The proofs are virtually identical, with the downward transition $d$ in place of the trying probability $a$. The key property is that the composition of downward kernels is downward, which is easy to check.



\subsection{Proof of \texorpdfstring{\cref{res:lower_bound}}{Lemma \ref{res:lower_bound}}} \label{sec:proof_lower_bound}

Fix $\th$ and $\th'$. For each $h$, supermultiplicativity (see \cref{res:most_discerning_authentication}) gives
\[
	\a (\th' | \th + h) \geq \a (\th' | \th) \a (\th | \th + h).
\]
Subtract $\a (\th'| \th)$ from each side to get
\begin{equation*}
\begin{aligned}
	\a (\th' |\th + h) - \a (\th' |\th) 
	&\geq \a (\th' | \th) (\a (\th | \th + h)  - 1) \\
	&= \a(\th' | \th) [\a(\th | \th + h)  - \a(\th| \th)].
\end{aligned}
\end{equation*}
Dividing by $h$ and passing to the limit as $h \downarrow 0$ and $h \uparrow 0$, we see that whenever $D_2 \a (\th' | \th)$ exists, we have 
\[
    -\l_+(\th) \a ( \th'| \th) \leq D_2 \a (\th' | \th) \leq \l_-(\th) \a ( \th'| \th).
\]

Since $\a$ satisfies \eqref{it:AC} and \eqref{it:integrable}, we can use absolute continuity to convert these local bounds into global bounds. Fix a report $\th'$. Define the function $\D$ on $[\ubar{\th}, \bar{\th}]$ by 
\[
	\D (\th) = \frac{ \a (\th' | \th)}{\L ( \th' | \th)}.
\]
By construction, $\D(\th') = 1$. We claim that $\D(\th) \geq 1$ for all $\th$. Since $\L(\th' | \th)$ is bounded away from $0$, the function $\D$ is absolutely continuous. Therefore, the functions $\D$, $\a(\th' | \cdot)$, and $\L(\th' | \cdot)$ are simultaneously differentiable almost everywhere. If $\th > \th'$ and these three functions are simultaneously differentiable at $\th$, we have
 \[
    \D'(\th)
    = \frac{1}{\L(\th' | \th)} \Brac{ D_2 \a(\th' |\th) + \l_+ (\th) \a ( \th' | \th)}
     \geq 0.
\]
If $\th < \th'$ and these three functions are simultaneously differentiable at $\th$, we have
\[
    \D' (\th)
    = \frac{1}{\L(\th' | \th)} \Brac{ D_{2} \a (\th' | \th) - \l_- (\th) \a(\th' | \th) } \leq 0.
\]
Since $\D$ is absolutely continuous, it follows from the fundamental theorem of calculus that $\D(\th) \geq \D(\th') = 1$ for all $\th$. 

\subsection{Universal measurability}  \label{sec:measurability}

We begin by introducing universal measurability. For a more detailed discussion with proofs, see \citet[Chapter 7]{BertsekasShreve}. Let $(X, \XX)$ be a measurable space. Given a probability measure $\mu$ on $(X, \XX)$, let $\overline{\XX}_{\mu}$ denote the $\mu$-completion of $\XX$, i.e., the $\s$-algebra generated by $\XX$ and all $\mu$-null sets of $\XX$. The universal completion of $\XX$, denoted $\overline{\XX}$, is the intersection $\cap_{\mu} \overline{\XX}_{\mu}$, where the intersection is taken over all probability measures $\mu$ on $(X, \XX)$. It can be shown that $\overline{\overline{\XX}} = \overline{\XX}$.

A function from $(X,\XX)$ to $(Y, \YY)$ is universally measurable if it is $(\overline{\XX}, \YY)$-measurable. It can be shown that $(\overline{\XX}, \YY)$-measurability is equivalent to $(\overline{\XX}, \overline{\YY})$-measurability. Similarly, it can be shown that any probability kernel from $(X, \overline{\XX})$ to $(Y, \YY)$ can be uniquely extended to a probability kernel from $(X, \overline{\XX})$ to $(Y, \overline{\YY})$. Given $(X, \XX)$ and $(Y, \YY)$, a probability kernel from $(X, \overline{\XX})$ to $(Y, \overline{\YY})$ is called
universally measurable. 

On a topological space $X$, the Borel $\s$-algebra is denoted by $\BB(X)$. For Polish spaces $X$ and $Y$, we have $\BB ( X \times Y) = \BB (X) \otimes \BB(Y)$. The left side is the $\s$-algebra generated by the product topology on $X \times Y$. The right side is the $\s$-algebra generated by all rectangles with Borel-measurable sides. 

Now we return to the model. We make the following standing technical assumptions. The sets $\Th$, $T$, and $X$ are Polish spaces. The function $\pi \colon T \times \Th \to \D(S)$ is Borel measurable (with $\D(S)$ viewed as a subset of $\R^{S}$).%
\footnote{We prove the measurability results in the nonbinary testing framework, which includes the main model as a special case.} In a mechanism, the message spaces $M$ and $M'$ are Polish, and all maps and probability kernels are universally measurable. Universally measurable sets are convenient because of the following measurable projection theorem \citep[Proposition 8.4.4,~p. 264]{Cohn2013}.

\begin{thm}[Measurable projection] \label{res:measurable_projection} 
Let $(X, \XX)$ be a measurable space, $Y$ a Polish space, and $C$ a set in the product $\s$-algebra $\XX \otimes \BB(Y)$. Then the projection of $C$ on $X$ belongs to $\overline{\XX}$.
\end{thm}

The definition of $\th$-discernment imposes an inequality for each type $\th'$. If there are uncountably many types, this can create measurability problems. Using the measurable projection theorem, we can show that the score conversion in the definition of $\tau \succeq_\th \psi$ can be selected in a universally measurable way.

\subsection{Measurable selection of score conversion} \label{sec:measurable_selection_score_conversion}

For each triple $(\th, \tau, \psi) \in \Th \times T^2$ such that $\tau \succeq_\th \psi$, there exists an associated score conversion satisfying \cref{def:discernment_order_general}.  Here we show that this score conversion can be selected in a universally measurable way.

We represent the space of increasing Markov transitions $k \colon S \to \D(S)$ as a polytope $\KK$ in $\R^{S \times S}$ consisting of vectors $k = (k(s' | s))_{s,s' \in S}$. Define the subset $G$ of $\Th \times T^2 \times \KK$ to consist of all tuples $(\th, \tau, \psi, k)$ such that $k$ satisfies the conditions in the definition of $\tau \succeq_\th \psi$. We will show below that $G$ is in $\overline{ \BB ( \Th \times T^2)} \otimes \BB(\KK)$. Then the projection of $G$ onto $\Th \times T^2$, which we call $D$, is in $\overline{\BB ( \Th \times T^2)}$ by the measurable projection theorem. The measurable projection theorem also guarantees that the section correspondence $(\th, \tau, \psi) \mapsto G_{\th, \tau, \psi}$ on $D$ is weakly measurable,\footnote{That is, the lower inverses of open sets are measurable. For each open subset $A$ of $\mathcal{K}$, the lower inverse of $A$ equals the projection of  $G \cap (\Th \times T^2 \times A)$ onto $\Th \times T^2$.}   where $D$ is endowed with the restriction of the $\s$-algebra $\overline{\BB ( \Th \times T^2)}$. Finally, this section correspondence has nonempty, closed values, so we apply the Kuratowski--Ryll-Nardzewski selection theorem \citep[18.13, p.~600]{AliprantisBorder2006} to get the desired universally measurable selection. 

Now we check that $G$ is in $\overline{ \BB ( \Th \times T^2)} \otimes \BB(\KK)$. On $\Th \times T^2 \times \KK$, define the real-valued functions $f_s$ for each $s$ in $S$, and $g_U$ for each upper set $U \subset S$ by
\begin{equation*}
\begin{aligned}
    f_s ( \th, \tau, \psi, k) &= (\pi_{\tau| \th} k) (s) - \pi_{\psi | \th} (s),\\
    g_U (\th, \tau, \psi, k) &= \sup_{\th'} \Brac{ (\pi_{\tau| \th'} k) (U)  - \pi_{\psi | \th'} (U)}.
\end{aligned}
\end{equation*}
The set $G$ is the intersection of $\cap_{s} [ f_s = 0]$ and $\cap_{U} [g_U \leq 0]$. Therefore, it suffices to check that these functions are all $(\overline{ \BB (\Th \times T^2)} \otimes \BB(\KK), \BB(\R))$-measurable. For each function $f_s$, this is implied by the Borel measurability of $\pi$. For each upper set $U$, we check that $g_U$ is a Carath\'{e}odory function. For each fixed $(\th, \tau, \psi)$, the function $g_U ( \th, \tau, \psi, \cdot)$ is continuous. For each fixed $k$, the function $g_U ( \cdot, k)$ is $(\overline{\BB (\Th \times T^2)}, \BB(\R))$-measurable because the term in brackets, viewed as a function of $(\th, \tau, \psi, \th')$ is $(\BB ( \Th \times T^2 \times \Th), \BB(\R))$-measurable. Hence, the supremum over $\th'$ is $(\overline{\BB ( \Th \times T^2)}, \BB(\R))$-measurable by the measurable projection theorem.\footnote{For any bounded function $f \colon X \times Y \to \R$, define $F \colon X \to \R$ by $F(x) = \sup_{y \in Y} f(x,y)$. For any real $t$, the preimage $[F > t]$ is the projection of the preimage $[f > t]$ onto $X$.} Therefore, $g_U$ is Carath\'{e}odory function. By \citet[4.51, p.~153]{AliprantisBorder2006}, $g_U$ is $(\overline{ \BB (\Th \times T^2)} \otimes \BB(\KK), \BB(\R))$-measurable.

\subsection{Measurable selection of more \texorpdfstring{$\th$-discerning}{theta-discerning} test}

Consider the setting of \cref{res:implementation_sharp}.  For each $(\psi, \th) \in \Th \times T$, there exists a test $\tau \in \hat{T}(\th)$ such that $\tau \succeq_\th \psi$. Here we show that this test can be selected in a universally measurable way.

Consider the following subsets of $\Th \times T^2$:
\[
   A = \{ (\th, \psi, \tau): \tau \succeq_\th \psi \}, \qquad
   B = \{ (\th, \psi, \tau): \tau \in \hat{T} ( \th) \}.
\]
It suffices to check that $A \cap B$ is in $\overline{\BB(\Th \times T)} \otimes \BB(T)$. Then the section correspondence $(\th, \psi) \mapsto (A \cap B)_{\th, \psi}$ has a $(\overline{\BB(\Th \times T)}, \BB(T))$-measurable selection by the same argument from \cref{sec:measurable_selection_score_conversion}. We check that $A$ and $B$ are each in $\overline{\BB(\Th \times T)} \otimes \BB(T)$. The set $\KK$ is compact, and by assumption $\pi$ is continuous, so it is straightforward to check that $A$ is closed. By assumption, $\hat{T}$ has closed values and is weakly $(\overline{\BB(\Th)}, \BB(T))$-measurable, so the correspondence $(\th, \psi) \mapsto \hat{T}(\th)$ has closed values and is weakly $(\overline{\BB(\Th \times T)}, \BB(T))$-measurable. Its graph, $B$, is therefore in $\overline{\BB(\Th \times T)} \otimes \BB(T)$ by \citet[18.6, p.~596]{AliprantisBorder2006}. 

\end{document}